\documentclass[a4paper,fleqn]{cas-dc}

\usepackage[authoryear]{natbib}
\usepackage{amsmath,amssymb,amsfonts}
\usepackage{algorithmic}
\usepackage{graphicx}
\usepackage{textcomp}
\usepackage{xcolor}
\usepackage{times}
\usepackage{soul}
\usepackage{url}
\usepackage{amsmath}
\usepackage{booktabs}
\usepackage{algorithm}
\usepackage{algorithmic}
\usepackage{amsthm}
\usepackage{subfig}
\newdefinition{definition}{Definition}

\usepackage{color}
\usepackage{soul}
\usepackage[dvipsnames]{xcolor}
\usepackage{soul}
\soulregister\cite7
\soulregister\section1
\soulregister\subsection1
\soulregister\ref7

\def\tsc#1{\csdef{#1}{\textsc{\lowercase{#1}}\xspace}}
\tsc{WGM}
\tsc{QE}
\tsc{EP}
\tsc{PMS}
\tsc{BEC}
\tsc{DE}

\begin{document}
\let\WriteBookmarks\relax
\def\floatpagepagefraction{1}
\def\textpagefraction{.001}
\shorttitle{Discovering Urban Functional Zones}
\shortauthors{Wen Tang et~al.}

\title [mode = title]{Discovering Urban Functional Zones from Biased and Sparse Points of Interests and Sparse Human Activities}



\author[1]{Wen Tang}[role=Student,orcid=0000-0003-0283-5493]
\cormark[1]
\ead{wtang6@ncsu.edu}


\address[1]{Department of Electrical and Computer Engineering, North Carolina State University, Raleigh, NC 27606, USA}

\author[2]{Alireza Chakeri}
\ead{chakeri@mail.usf.edu}
\address[2]{Valassis Digital, 3020 Carrington Mill Blvd, Morrisville, NC 27560, USA}

\author[1]{Hamid Krim}
\ead{ahk@ncsu.edu}




\cortext[cor1]{Corresponding author}


\begin{abstract}
With rapid development of socio-economics, the task of discovering functional zones becomes critical to better understand the interactions between social activities and spatial locations.  {In this paper, we propose a framework to discover the real functional zones from the biased and extremely sparse Point of Interests (POIs). To cope with the bias and sparsity of POIs, the unbiased inner influences between spatial locations and human activities are introduced to learn a balanced and dense latent region representation. In addition, a spatial location based clustering method is also included to
enrich the spatial information for latent region representation and enhance the region functionality consistency for the fine-grained region segmentation.}
 {Moreover, to properly annotate the various and fine-grained region functionalities,} we estimate the functionality of the regions and rank them by the differences between the normalized POI distributions  {to reduce the inconsistency caused by the fine-grained segmentation.}  {Thus, our whole} framework is able to properly address the biased categories in sparse POI data and explore the true functional zones  {with a fine-grained level.} To validate the proposed framework, a case study is evaluated by using very large real-world users GPS and POIs data from city of Raleigh. The results demonstrate that the proposed framework can better identify functional zones than the benchmarks, and, therefore, enhance understanding of urban structures with a finer granularity under practical conditions.
\end{abstract}



\begin{keywords}
Functional Zones discovering \sep Latent region representation learning \sep Sparse and bias POIs \sep GPS data, Conditional Random Filed Clustering \sep Function Annotation
\end{keywords}

\ExplSyntaxOn
\keys_set:nn { stm / mktitle } { nologo }
\ExplSyntaxOff

\maketitle

\section{Introduction}

The development of the network infrastructures and cellphone hardwares led to massive amount of mobile data with geographical information. In order to leverage the knowledge of such huge mobile data, it is critical to develop a digital map and its functional zones. Functional zones not only help individuals to understand a complex city, but also suggest a scientific urban planning. Additionally, functional zones are also useful for business sites management and tourism pattern discoveries, since they reveal the intention of human activity trajectories. It also answers the key questions of where and why people go visiting. On the other hand, the urbanization and modern civilization recently occur at an unprecedented rate. Hence, the task of functional zones discovery becomes essential in order to understand the interactions between social characters and spatial locations.

 {To address this problem, two main strategies have recently been developed in the literature.} One is to reveal and visualize the complex spatial interactions between different functional zones \cite{guo2009flow,yan2009visual}, where any spatial interaction is quantified on the basis of the trips associated with a location and a timestamp of origin / destination. Zhang et al. \cite{zhong2014detecting} defined a weighted network, where nodes are urban areas, edges are travel records and their weights are the number of trips. They then detected the graph centralities and clustered graph community structures to obtain the hubs, centers and borders. In \cite{liu2015revealing,demsar2014edge}, the authors used community detection methods to identify the functional regions. Peng et al. \cite{peng2012collective} applied Non-negative Matrix Factorization (NMF) to estimate different patterns and clustered them into areas. However, the quantitative analysis of spatial interactions did not carry enough information to explore the functional zones.

\begin{figure*}[htbp]
	\centerline{\includegraphics[width=0.8\textwidth]{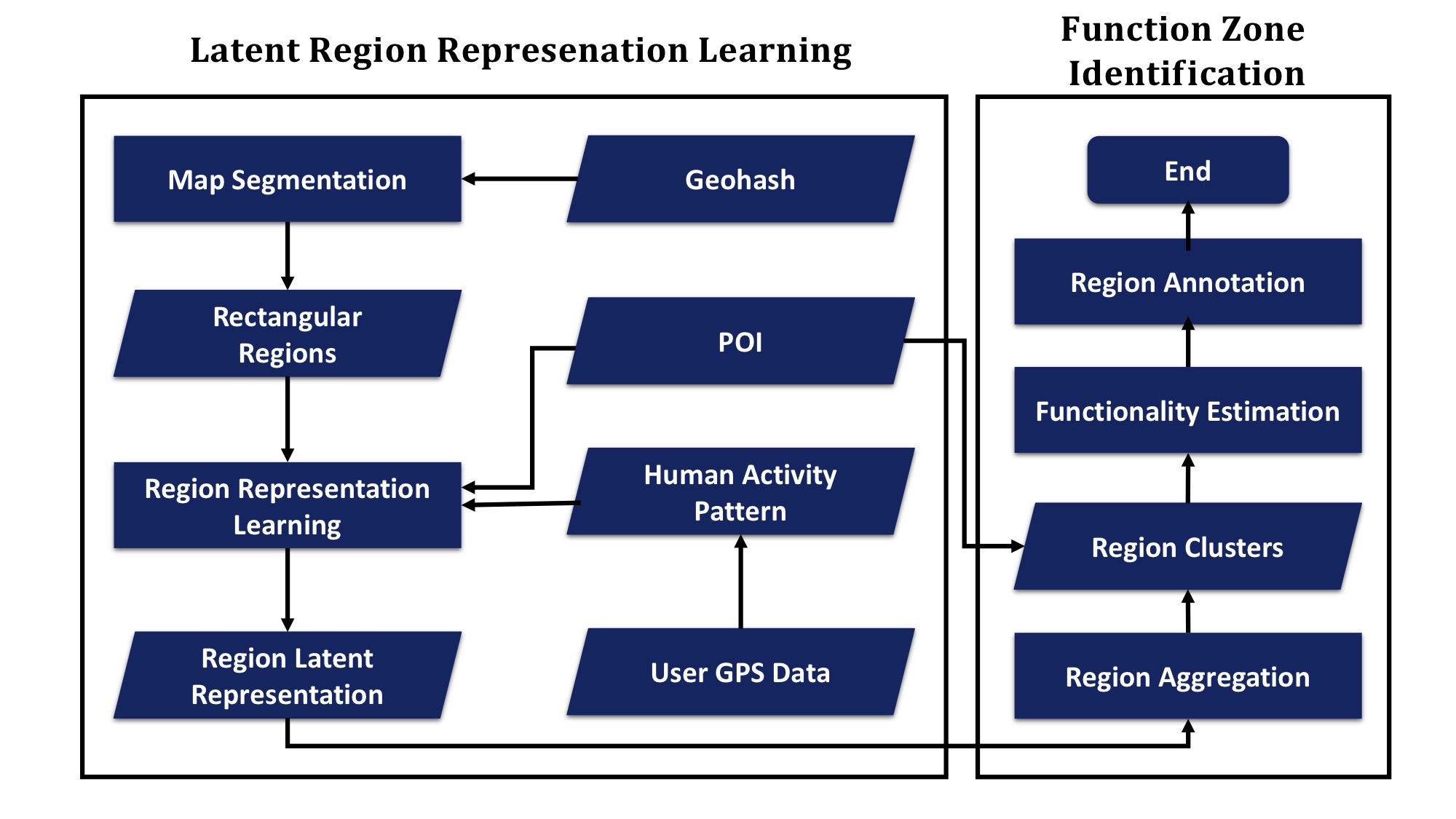}}
	\caption{Framework of Functional Zones Revealing.}
	\label{fig:framework}
\end{figure*}

Another strategy is to introduce POIs to jointly combine them with the spatial interactions in order to discover functional zones \cite{gao2017extracting, yuan2015discovering, wang2018using, liu2015revealing,yuan2018discovering}. The authors in \cite{yuan2015discovering, gao2017extracting, yuan2018discovering} applied the Latent Dirichlet allocation (LDA) based topic model to POIs and human activity behaviors, such as, user check-in data and taxi trajectories. \cite{han2015discovering} clustered the bus smart card data and identified different functional zones using POIs. \cite{wang2018using} used NMF method in \cite{boarnet2001influence} to identify functional regions based on POI data and taxi trajectory data. They then quantitatively analyze the spatio-temporal information for annotation.

However, all above methods neither leverage a real intrinsic structure between POIs and human activity patterns, nor deal with extremely sparse and biased real-world data. Additionally, the spatial information is also failed to be considered when the regions are aggregated. Generally, the methods in the literature assumed that the data follow some fixed probability distributions. However, our real-world data is highly sparse and less than $1\%$ of the regions carry POI information. This is insufficient to estimate such parameters of the distributions. Also, compared to their POIs where they are generally unbiased and contain all different categories, in our dataset, only the commercial related POIs are provided. In order to cope with such biased and sparse information, we propose a novel functional zone identification framework to fuse the information between POI and human activity patterns. Instead of using probability assumptions, we introduce and transfer an intrinsic interaction structure between POIs and human activity patterns that was discovered in \cite{wang2017human} with unbiased data to bring the extra knowledge to alleviate the biased effects, and cooperates with matrix factorization to address the high sparsity, without any non-negative constraints. Moreover, the prior spatial knowledge is also used for region aggregation to confront against the bias issue. A novel optimization method is also proposed to learn the latent regions representation based on such intrinsic interaction structure, which is applicable to biased/unbiased and sparse/non-sparse data. 
To reduce the bias influence, CRF is applied to functional zones clustering to preserve both spatial information and their discriminative representations. The difference of normalized POI is then computed for functional zones annotation to emphasize the significance of different functionalities. The very large real-world users GPS and POIs data of city of Raleigh are used to evaluate our framework and demonstrate a better functional zones identification and a finer granularity than the benchmarks. The proposed framework is summarized in Figure. \ref{fig:framework}.



Our main contributions can be summarized as:
\begin{itemize}
    \item [\textbf{(1)}] A novel latent region representation learning method is proposed based on intrinsic interaction of unbiased POIs and human activity patterns to overcome the biased data.
    \item [\textbf{(2)}] To exploit the prior spatial information and alleviate both bias effects and the imbalance POIs of map segmentation, CRF is used as an unsupervised clustering method, which is regularized by the spatial relationship.
    \item [\textbf{(3)}] A novel and efficient functional zone detection framework is proposed for both biased/unbiased and sparse/non-sparse data. This achieves a better functional zones identification and a finer granularity than the benchmarks based on real data.
\end{itemize}

The rest of the paper is organized as follows: Section \ref{sec:latentlearing} introduces the preliminaries and the details on the latent region representation learning procedure. Functional regions aggregation and identification methods based on function estimation are discussed in Section \ref{sec:identification}. The results are shown in Section \ref{sec:experiment}.  {And more discussion of properly using our method is discussed in Section \ref{sec:discussion}.} Finally, Section \ref{sec:conclusion} concludes the paper.

\section{Latent Region Representation Learning} \label{sec:latentlearing}
In order to discover different functional zones, we segment our city into small rectangle regions, which are considered as the basic region units. Based on these region units, a novel algorithm is proposed to learn latent semantic representation of each region unit that captures the interaction information between POIs and human activities.

\subsection{Map Segmentation}
\begin{figure}[htbp]
	\centerline{\includegraphics[width=0.44\textwidth]{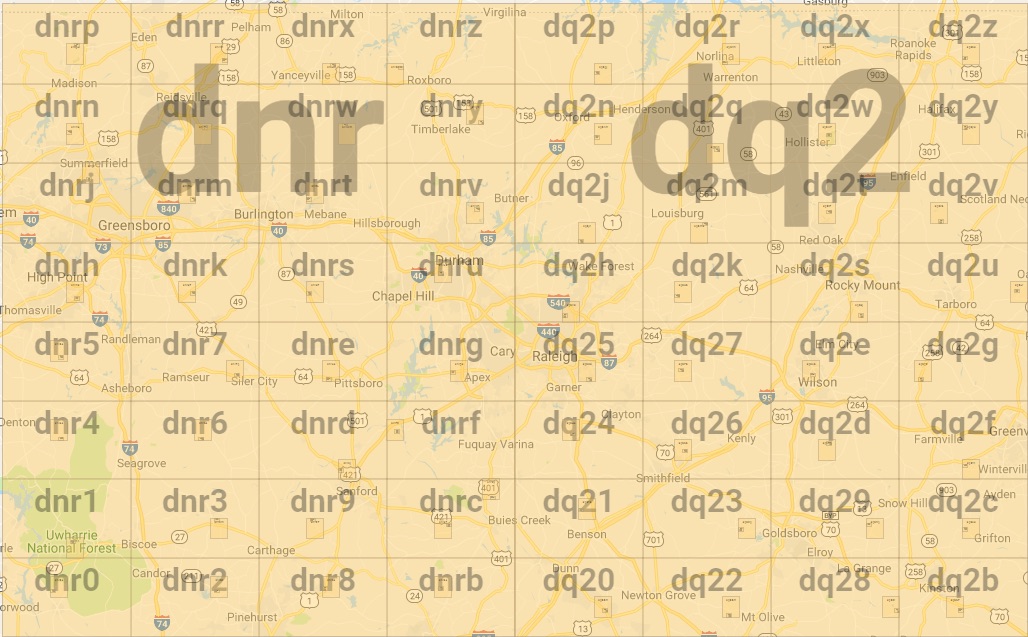}}
	\caption{Examples of Geohash level $3$ and Geohash $4$ Segmentation. $2$ big Geohash level $3$ regions cover $64$ Geohash level $4$ regions. }
	\label{fig:geohash}
\end{figure}

To speed up the map segmentation procedure, the well-known Geohash geocoding system is adopted to segment the map into equal-sized regions. geohashes express the girded regions by short alphanumeric strings. The greater precision is obtained with a longer string and a larger Geohash level. For example, two $156.5 km \times 156 km$ rectangle areas of Geohash level $3$ can perfectly cover the city of Raleigh, whereas it can also be accurately represented by 64 $39.1 km \times 19.5 km$ regions of Geohash level $4$ as shown in Figure  \ref{fig:geohash}. In total, there are 12 levels of Geohashes to present different area granularity. Geohash level 6 is used in our experiments.

\subsection{Region Feature Learning}

\subsubsection{Preliminaries}
\begin{definition}
	Trajectory: A trajectory of a user is a sequence of time stamped location points, i.e.,$Traj=(lp_1,\dots,lp_n)$, where $lp_i=(lat,lon,t),$ represents the latitude and longitude at time $t_i$, $ \forall i=1,\dots n$.
\end{definition}

\begin{definition}
	Distance and Time Period: The geospatial distance between two points $lp_i$ and $lp_j$ is defined as $D(lp_i,lp_j)$. Time period $TP(lp_i,lp_j)=|lp_i.t-lp_j.t|$ is the time interval between two points.
\end{definition}

\begin{definition}
	Human Activity: A human activity $h_i$ of a user is defined such that the user stays in a geographical region within a given $Dr$ distance over a given time period $Tr$. $h^j_i$ is decided by a set of consecutive points $L=\{lp_k,lp_{k+1},\dots,lp_m\}$. In particular, when for all $ k<i\leq m$, $D(lp_k,lp_i)\leq Dr$, $D(lp_k,lp_{m+1})>Dr$ and $TP(lp_k,lp_i)\leq Tr$, $h_i=(lat,lon,t_a,t_l),$ where $t_a=lp_k.t$ is the arrival time and $t_l=lp_m.t$ is the corresponding leaving time. Also, $h_i.lat=\frac{1}{|L|} \sum_{i=k}^m lp_i.lat$ and $h_i.lon=\frac{1}{|L|} \sum_{i=k}^m lp_i.lon$ are respectively the latitude and longitude of the human activity geospatial location.
\end{definition}

\begin{definition}
	Human Activity Trajectory: A human activity trajectory is a sequence of human activities, i.e. $HT=\{h_1 \to h_2 \dots \to h_q\}$ where $h_i$ is a human activity.
\end{definition}

\begin{definition}
	Human Activity Information: A human activity information $A$ is a triplet consisting of the origin region $R_o$, the destination region $R_d$ and departure time $t_l$ or arrival time $t_a$. Leaving human activity information is defined as $A_L=(R_o,R_d,t_l)$, and arriving human activity information is then defined as $A_A=(R_o,R_d,t_a)$. For each human activity $h_i$ in the human activity trajectory $HT$, a map $f$ has been used to map the geospatial location to different Geohash regions, i.e., $R_i=f(h_i.lat,h_i.lon)$. The leaving and arriving time are respectively obtained by $t_l=h_{i}.t_l$ and $t_a=h_{i}.t_a$.
\end{definition}


\begin{definition}
    Bag-of-POIs: Let assume that there are $p$ different POI categories, and $p_j$ shows the number of the $j$th POI category in the map. Then, the bag-of-POIs for the $i$th region unit is defined as $poi_i = (p_1^i,\dots,p_p^i), ~\forall i=1,\dots,p.$
\end{definition}
Since the bag-of-POIs uniquely represents different functional zones, it can be used to represent each region to reserve the functional information and obtain the POI information matrix $P$ as in the following:
\begin{definition}
    POI Information Matrix: Let each column of this matrix be the Bag-of-POIs of $i^{th}$ region unit, i.e.
\[P=
\begin{bmatrix}
poi_1^T
poi_2^T
\dots
poi_p^T
\end{bmatrix}.
\]
\end{definition}
\begin{figure*}[htbp]
	\centerline{\includegraphics[width=0.7\textwidth]{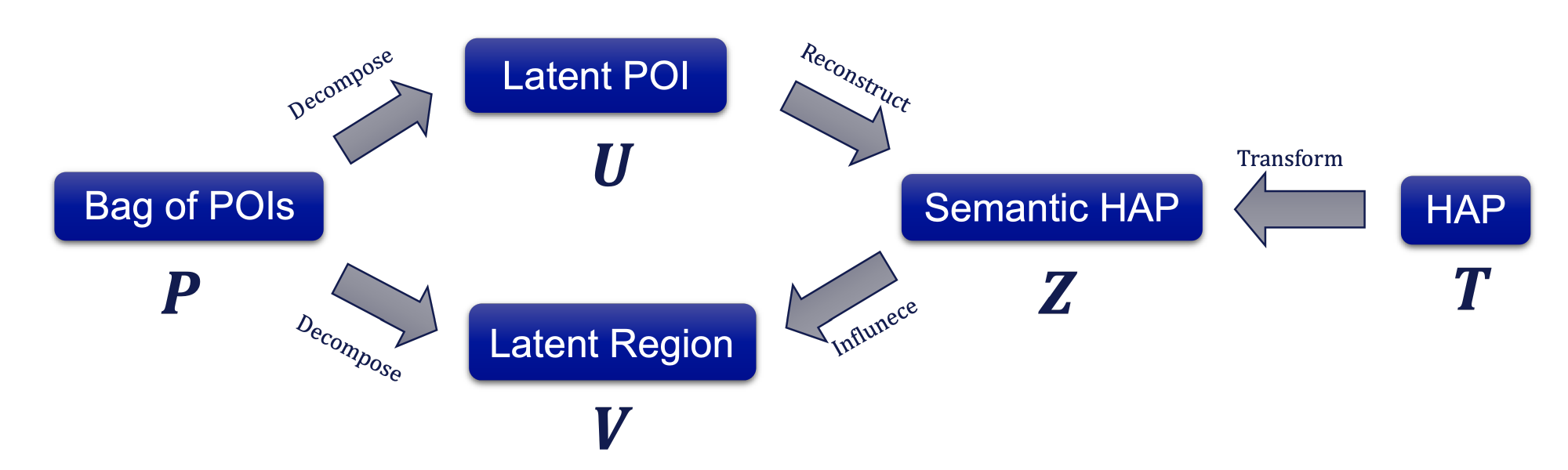}}
	\caption{Latent Region Representation Learning Model.}
	\label{fig:latentrepresentionlearning}
\end{figure*}
\begin{definition}
Human Activity Pattern (HAP) Matrix: Based on each human activity information $A_L=(R_o, R_d, t_l)$ and $A_A=(R_o, R_d, t_a)$, let each element of matrix $C_k = [c_{ij}^k]$ represents the number of $A_L$ (or $A_A$) from $i^{th}$ region unit to $j^{th}$ region unit, during the time period $k$. Then we define the HAP matrix as
\[T=
\begin{bmatrix}
C_1,C_2,\dots,C_s
\end{bmatrix}
^T
\]

\noindent Each time period is defined as one hour. 24 time periods are considered in one day. That is, $t_l \in $ [0:00,1:00) $,\dots, t_l \in $ [23:00,24:00). In our experiment, both $T$ based on $A_L$ and $A_A$ are concatenated together.
\end{definition}

\subsubsection{Latent Region Representation Learning}
In order to discover the functional zones, we need to explore the discriminative representations of various functional zones. One useful functional zone related information is POIs, as different POIs (such as fast food, coffee bar, shopping mall) attract individuals for different purposes. In addition, more POIs translates into higher probabilities for their corresponding functionalities. As a result, Bag-of-POIs were  {first} introduced in our framework  {in a naive way}.

However, POIs are not only the determining factor of the region functionalities. The region functionality is also decided by the intention of individuals trajectories. \cite{chen2015social} discovered that various region functional zones have their corresponding HAPs varies in different time periods. This implies that HAPs which reflect semantic meanings of Human intentions, another important factor for functional zones.  {HAPs is therefore, also included into our framework to enrich the knowledge of the region functional zones discovery.}

 {In contrast to the conventional methods that proposed on the balanced and sufficient datasets, when the POIs data is very sparse, it is insufficient for small number of biased POIs to precisely estimate or learn the parameters of assumed distributions in  \cite{yuan2015discovering,gao2017extracting,yuan2018discovering} and the variables in \cite{wang2018using, liu2015revealing}.} We hence  {decompose the POI and region features into latent dense representations and} propose an intrinsic interaction between  {the latent} POIs and HAPs in our method.  {The intrinsic interaction that we proposed is inspired from \cite{wang2017human}, as they revealed that } $80\%$ of HAPs can be successfully predicted by POI information.  {That is ,} POIs information can almost entirely reconstruct HAPs.  {Since }this discovery was  {also fully} data-driven and leveraged on the unbiased data, such intrinsic knowledge  {can be} directly transferred into our framework to cope with the POI bias. Additionally, since region functionalities are critical to affect human travel behaviors and infers traffic pattern, a transformation between region representation and HAPs should be  {also incorporated.}

Consequently, as shown in Fig. \ref{fig:latentrepresentionlearning}, the POI information matrix $P$ is decomposed into two latent and dense spaces $U$ and $V$. And HAP matrix $T$ is transformed by matrix $Q$ into a semantic latent space $Z$ to reduce the dimensions and preserve useful information. According to the transferred intrinsic knowledge leveraged from unbiased data, the transformed HAP $Z$ is also reconstructed by some important latent POI features $U$ that are gleaned from the original POI information in a latent space. The coefficients matrix $A$ that regularized by $L_1$ norm are used to select the important latent POI features in $U$ for each region. Meanwhile, the latent HAP representation $Z$ also influences the latent region representation $V$ that is also derived from POI information in the same latent space. Besides, to avoid the lack of complete POI information, an indicator mask is also used to deal with the sparsity of POI matrix.

Hence, our final information fusion method for the latent region representation learning is formulated as follows:
\begin{equation}\label{equ:main}
\begin{split}
\arg\min_{\substack{U,V,Q\\W,A,Z}} &~\frac{1}{2} \|I\circ
(P-UV)\|_F^2+\frac{\lambda_1}{2} \|QT-Z\|_F^2\\
&+\frac{\lambda_2}{2} \|Z-U^TA\|_F^2 + \lambda_3  \|A\|_1+\frac{\lambda_4}{2} \|V-WZ\|_F^2\\
&+\frac{\lambda_5}{2} (\|U\|_F^2+\|V\|_F^2+\|Q\|_F^2+\|W\|_F^2),
\end{split}
\end{equation}
where $I$ is a indicator mask. $I_{i,j}=1$ for non-empty entries and $I_{i,j}=0$ for others, $\circ$ is an element-wise product.  $\lambda_1,\lambda_2,\lambda_3,\lambda_4, \text{ and }\lambda_5$ are tuning parameters. $P\in \mathbb{R}^{p\times r}$ is decomposed into two latent semantic matrices $U\in \mathbb{R}^{p\times k}$ and $V\in \mathbb{R}^{k\times r}$. $T\in \mathbb{R}^{q\times r}$ is transformed into its latent transition pattern matrix $Z\in \mathbb{R}^{k\times r}$ by a transform matrix $Q\in \mathbb{R}^{k\times q}$. The latent transition pattern $Z$ receives the information from the latent POI features by using sparse coefficients $A \in \mathbb{R}^{p\times r}$ to choose a specific POI in $U$, and propagates the information back to the latent region feature $V$ by $W \in \mathbb{R}^{k\times k}$ regression. The last term in Equation (\ref{equ:main}) is the regularization for avoiding overfitting.

We, therefore, alternatively search one variable when others are fixed. The updating rule of each variable based on gradient descent are also summarized in Algorithm \ref{alg1},  {where $\tau(\cdot)$ is the element-wise soft thresholding operator, a solution of $L_1$-norm and is used in many sparse learning \cite{fista,tang2016analysis,tang2018structured,tang2019convolution,tang2019analysis,tang2020deep,tang2021deep}.}

\begin{algorithm} 
	\caption{Latent Region Representation Learning} 
	\label{alg1} 
	\begin{algorithmic} 
		\REQUIRE POI matrix $P$, HAP matrix $T$, and hyper-parameter $\lambda_1,\lambda_2,\lambda_3,\lambda_4, \text{ and } \lambda_5$
		\ENSURE Latent POI feature $U$, latent Region representation $V$, Sparse Coefficients $A$, transform matrix $Q$, latent transition pattern matrix $Z$, and regression matrix $W$
		\STATE t=1;
		\WHILE {$t<maxIteration$ and $L_t-L_{t+1}>\varepsilon$}
		\STATE \resizebox{0.9\columnwidth}{!}{$U_{t+1}=U_t-\alpha(-I\circ(P-UV)V^T-\lambda_2A(Z-U^TA)^T+\lambda_5U)$}
		\STATE \resizebox{0.9\columnwidth}{!}{$V_{t+1}=V_t-\alpha(-U^TI\circ(P-UV)-\lambda_4(V-WZ)+\lambda_5V)$}
		\STATE \resizebox{0.55\columnwidth}{!}{$Q_{t+1}=Q_t-\alpha(\lambda_1(QT-Z)T^T+\lambda_5Q)$}
		\STATE \resizebox{0.9\columnwidth}{!}{$Z_{t+1}=Z_t-\alpha(-\lambda_1(QT-Z)+\lambda_2(Z-U^TA)-\lambda_4)W^T(V-WZ)$}
		\STATE \resizebox{0.58\columnwidth}{!}{$A_{t+1}=\tau_{\alpha\lambda_3}\left(A_t-\alpha(-\lambda_2U(Z-U^TA))\right)$}
		\STATE \resizebox{0.58\columnwidth}{!}{$W_{t+1}=W_t-\alpha(-\lambda_4(V-WZ)Z^T+\lambda_5W)$}
		\STATE $\alpha=\rho\alpha$ \% $\rho$ is a decay factor
		\ENDWHILE
		\RETURN $U,V,Q,Z,A,W$
	\end{algorithmic}
\end{algorithm}

After solving the algorithm, the latent region representation $V$ comprehensively fuses both information of POI and HAP and the transferred knowledge. The latent transition pattern matrix $Z$ also contains the semantic meanings of HAP.  {Given $U,V,Q,Z,A,W$, we then can identify the functional zones based on the latent region representation $V$, a $k$-dimensional feature $(v_1^r,\dots,v_k^r)$ for each region unit $r$.}

\section{Functional Zone Identification} \label{sec:identification}
\subsection{Region Aggregation}
 {As the ground truth of region labels are not available at this moment, clustering is then naturally used to perform it to aggregate the regions.} Regions in the same clusters should have similar functionalities,  {while} those in different clusters  {regions should} provide various  {different} functionalities.
$K$-means clustering method is a naive and usual way to cluster region units to functional zones. However, $K$-means does not consider the spatial information of each region. When the segmentation grids is very small, a big place may segmented into different region units and carrying different number of POIs.  {Then the fine-grained region representations may be totally different from each other and lead to a worse clustering result by $K$-means.} For instance, a shopping mall is segmented into 3 region units. One of them carry the majority of POIs of different stores, while another two of them only have a few stores. There is then a huge difference among POI features of those region units.  {Then these 3 region units are more likely clustered into 2 various clusters rather than be clustered into the same cluster.}
 {Therefore, if the spatial information is not leveraged, $K$-means clustering based on the region representation is not enough for the finer granularity. When the segmentation is fine-grained, many functional zones are partitioned into different small parts with different kinds of POIs and various numbers of POIs, thus introducing more variance into the same functional zones.} In order to confront against such imbalance that caused by the fine-grids, we take the spatial location of each region units into account.  {In usual, when the granularity is fine, the 8 neighbor region units that surround this region unit should be more probably share the same cluster}. Hence,  {the spatial location information should also be utilized in the clustering as well to improve the consistency of the clustered region units. As the region representation and neighbor clusters will both influence the current region clustered result, the current region unit representation and their neighbours clusters are connected as a small random field that conditional independent with other region units.} Conditional Random Field (CRF) method is  {thus involved} to cluster the region units,  {because} CRF method aims to maximize a posterior (MAP) inference over each region unit  {representations and meanwhile also to }maximize  {each} cluster agreement between similar region units  {and neighbors} to induce more smoothness \cite{krahenbuhl2011efficient}. Each region unit can be regarded as a $k$-dimensional data point that is generated by a probability distribution, such as Gaussian. And a  {cluster} agreement penalty is also jointly used to calculate the  {cluster} generation probability of different region units. In our paper, the Gaussian distribution is regarded as the generative distribution of each region unit, and only $8$ neighbors of each region units are used in CRF to penalize the  {cluster} agreement. The process of CRF can be formulated as follows:

	\noindent(1) Let functional Zone $\mathcal{L}=\{l_1,\dots,l_c\}$ have $c$ different functional zones, and random variable $Y=\{y_1,\dots,y_r\}$ denote region-unit-level  {clustering}, such as $\{ cluster_1,\dots,$ $cluster_r\}$ that are un-annotated and finally assigned to each region unit, and random variable $R=\{r_1,\dots,r_r\}$ denote a feature of each region. In our experiment, the random variable $R$ is the latent region feature $V$. Then the conditional probability of  {cluster} $Y=l_K$ is assumed to follow a normal distribution:
	\begin{equation}
	P(Y=l_K|R)=N(\mu_{l_K},\Sigma_{l_K}),
	\end{equation}
	\noindent(2) For each region unit $R_i$, the joint probability of  {cluster} $y_i\in \mathcal{L}^r$ in CRF is given by:
	\begin{equation}
	\begin{split}
	&P(y_i,y_j)=exp\{-\sum_{(i,j)\in N} V(i,j)\},\\
	&V(i,j)=\beta\delta(y_i,y_j)=\begin{cases}
	\beta & \text{if } y_i=y_j\\
	-\beta & \text{else}\\
	\end{cases},
	\end{split}
	\end{equation}
	where $N$ is the neighbor set of region $R_i$.
	\noindent(3) The final target function is formulated as follows:
	\begin{equation}\label{equ:crf}
	\arg\max_{y\in \mathcal{L}^r}~P(r,y)=\frac{1}{F}\Pi_i P(y_i|r_i)\Pi_{i,j} P(y_i,y_j),
	\end{equation}
	where $F$ is a normalization term.

The join probability in Eq.(\ref{equ:crf}) can be solved by different approaches, such as Expectation-maximization (EM) algorithm  {in  \cite{dempster1977maximum}.}

\subsection{Functionality Estimation}
 {Due to the lack of the labels, after region aggregations,} functionality estimation of each cluster {is then another necessary step before annotating the} functional zones. {Once the estimated functionalities are measured, then the functional zones can be annotated based on these estimations. For the functionality estimation,} the  {unified} POI distribution  {of each cluster} is supposed to be adequate to carry the functionality.
However, the actual number of POIs is not the same in each region units. And the number of region units in each cluster is also different. Thus, to emphasize the significant functionality of different POIs in each cluster, the maximum normalized POI distributions is adopted for comparison, which reflects the importance proportion of contributions of different POIs. For the $s^{th}$ cluster, the vector of POI distribution is hence formulated as follows:

\begin{equation}
{PR^s}=(\frac{\sum_{i\in S} p^i_1}{|S|},\dots,\frac{\sum_{i\in S} p^i_p}{|S|}),
\end{equation}
where $S$ is a set of region units whose labels of the functional zone are $l_s$, and $|S|$ is the number of elements in the set $S$.
The normalized distribution then is:
\begin{equation}
NPR^s=\frac{PR^s}{\max(PR^s)}.
\end{equation}

\subsection{Region Annotation}
 {Finally, }region annotation is to associate the real functionalities with clusters. However, it is far from enough to name a functional zone by only using the frequency of POIs. Because many functionalities may appear in the same cluster, which leads to a hard region annotation. In our method, we use the difference of functionality to assign some semantic terms to each functional zone to understand its real function. The following conditions is utilized to annotate different functional zones:
\begin{itemize}
	\item[(1)]
	\noindent\textbf{
	The difference in the normalized POI distribution.} The difference $G_s$ of $s^{th}$ functional zone is calculated as follows,
	\begin{equation}
	G_s=(|\mathcal{L}|-1)NPR^s-\sum_{\substack{j\in \mathcal{L}, j\neq s}}NPR^j,
	\end{equation}
	where $|\mathcal{L}|$ is the number of functional zones. The positive elements of $G_s$ is the significant POIs in the functional zone $s$.
	\item [(2)]
    \noindent \textbf{
	Land-use maps of government.} Compared with land use or zoning map, it is more accurate for us to understand different functional zones we obtained. For example, several big industrial regions can help us fast locate such functional zones.
	\item[(3)]
	\noindent \textbf{
	Human-labelling.} As some information may lack in the POI information, people know more about well-known or common places that never show up in the POI information, such as, the label of universities, which is missing in our experiments.
\end{itemize}

\begin{figure}[!tbp]
	\centering
	\subfloat{\includegraphics[width=0.45\textwidth]{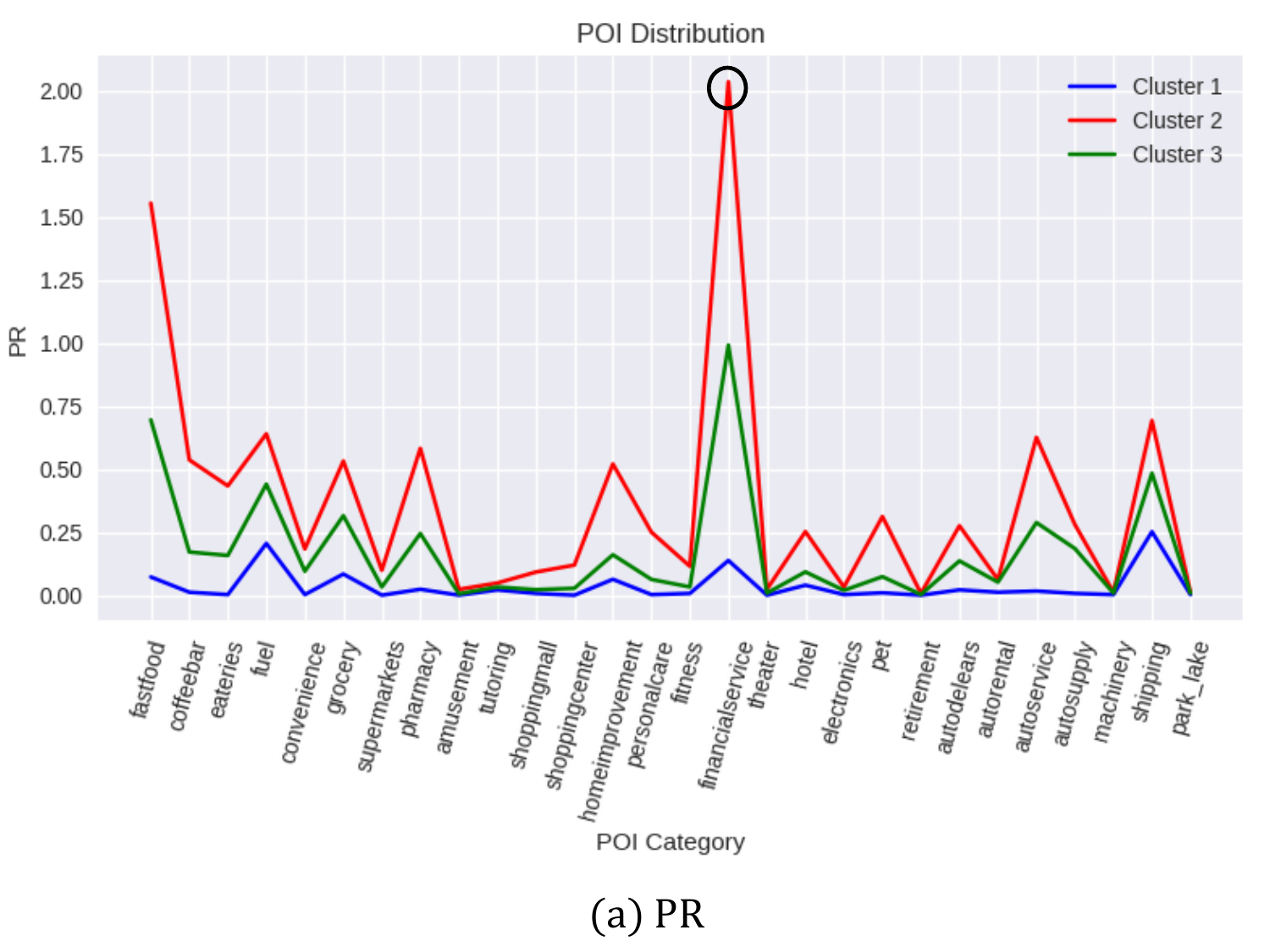}}
	\vfill
	\subfloat{\includegraphics[width=0.45\textwidth]{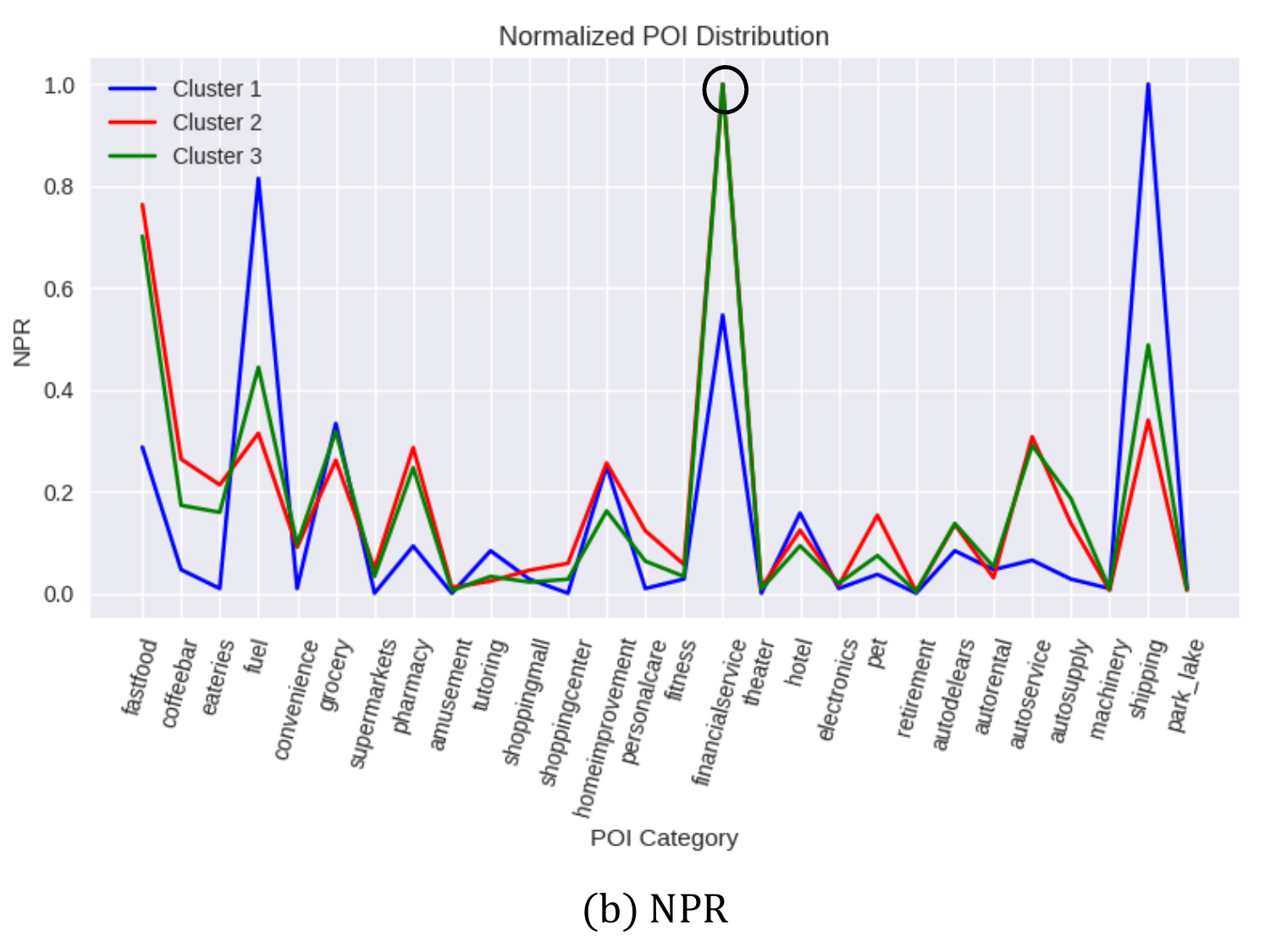}}
	\vfill
	\subfloat{\includegraphics[width=0.45\textwidth]{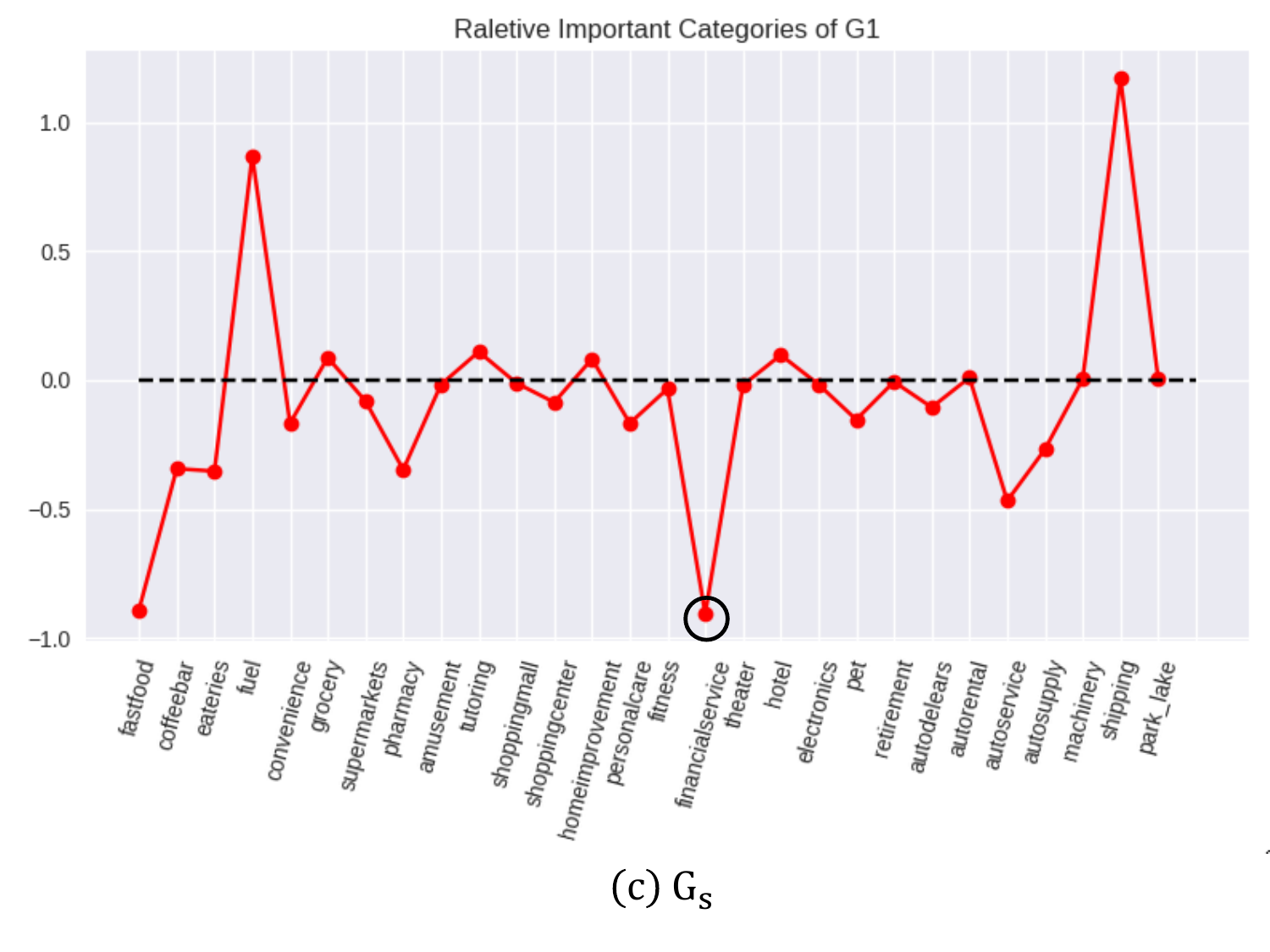}}
	\caption{Example of $PR^s$, $NPR^s$ and $G_s$.}
	\label{fig:pr-npr-g}
\end{figure}

All the examples of the metrics of the functionality estimation and the region annotation are shown in Fig. \ref{fig:pr-npr-g}.  {As shown in Fig. \ref{fig:pr-npr-g}, it is easy to show that the magnitude of the significant functionalities changed for different clusters after normalization by comparing (a) PR and (b) NPR. To show the role of (c) $G_s$, the example of $G_1$ can directly show its improvement of significance of important POIs. In (c), after calculating the difference of $G_1$, and only $G_1 >0$ is considered for annotation, it is clear that financial service (the point in the black circle) is not so significant for Cluster 1 as before, when comparing with Cluster 2 and Cluster 3. Because, although there are many financial services in Cluster 1, it also appears many times in Cluster 2 and Cluster 3. Thus actually, it is not distinguished among these clusters. Therefore, the difference $G_s$ could be used to improve the significance and distinction of the POIs for functionality annotation.}

\section{Experiments and Results}\label{sec:experiment}

\subsection{Experiment Settings}
Our experiments are evaluated on the following datasets:
\begin{itemize}
	\item[(1)]
	 Raleigh POI dataset is updated starting 2014. The category of POI has be listed in Table \ref{tab:poi}. There are 6,725 POIs and 1,233 regions with POI information. The sparsity of POI matrix $P$ is $99.11$\%.
	\begin{table}[htpb]
		\centering
		\resizebox{0.99\columnwidth}{!}{
		\begin{tabular}{cc|cc}
			\hline
			\textbf{Index} &\textbf{ POI Category }& \textbf{Index} & \textbf{POI Category}\\
			\hline
			1 &  fast food & 15& fitness\\
			2 & coffee bar & 16& financial service\\
			3& eateries & 17&theater\\
			4& fuel & 18&hotel\\
			5& convenience store & 19&electronics store\\
			6& grocery & 20&pets/veterinary\\
			7& supermarkets & 21& retirement\\
			8& pharmacy & 22& auto dealers\\
			9& amusement & 23& auto rental\\
			10& tutoring school & 24& auto service\\
			11& shopping mall & 25&auto supply\\
			12& shopping center & 26& machinery\\
			13& home improvement & 27& shipping store\\
			14& personal care & 28& park/lake(camping site)\\
			\hline
		\end{tabular}}
		\caption{POI Category}
		\label{tab:poi}
	\end{table}
	\item[(2)]
	\noindent User GPS dataset is collected during the workdays of March $15^{th}$ to May $20^{th}$ in 2018.  In this dataset, there are $1,026,726$ users GPS records that located in Raleigh. They are then transformed to 27,930,027 Human Activity Information. The number of regions with HAP is 16,321. The sparsity of HAP matrix $T$ is $99.97\%$.
	\item[(3)]
	\noindent Region Segmentation: We transformed Raleigh Map in to $1200~m~\times~ 609.4~m$ small rectangle regions that coding by Geohash level $6$. We totally obtain 16,384 segmented regions.
\end{itemize}
The statistics of all three datasets are listed in Table \ref{tab:stat}.
		\begin{table}[htpb]
		\centering
		\resizebox{0.98\columnwidth}{!}{
		\begin{tabular}{l l}
			\hline
			\textbf{POI Dataset} & \textbf{Statistics} \\
			\hline
			\#POI &  6,725\\
			\#Regions with POIs / \#Segmented Regions & 1,233/16,384\\
			Sparsity of POI Matrix& 99.11\%\\
			\hline
			\textbf{User GPS Dataset} & \textbf{Statistics}\\
			\hline
			\#Users &  1,026,726\\
			\#HAP &  27,930,027\\
			\#Regions with HAP / \#Segmented Regions & 16,321/16,384\\
			Sparsity of HAP Matrix& 99.97\%\\
			\hline
		\end{tabular}}
		\caption{POI Statistics}
		\label{tab:stat}
	\end{table}

\subsection{Benchmarks}
In our experiments, the dimension of the latent region representation is set to 10.

 {In order to avoid over-weighted the common but unimportant POIs, TF-IDF is applied to the POI matrix to adjust POIs. For example, people may go to some shopping center because of its cinema there, but they may go for dinner first. And the number of restaurants usually is much higher than the number of cinema. The region functionality is then apparently dominated by the restaurants if only based on the frequency of POIs. However, the true region functionality maybe the cinema. To avoid such over-weighted issues, TF-IDF is then used to adjust and balance the frequency and distinguish of POIs.} TF-IDF is applied by following the settings in \cite{yuan2015discovering,wang2017human}. We  {then }compared our results with the following benchmarks, and all benchmark methods build up on the POI matrix prepossessed by TF-IDF:
\begin{itemize}
  \item [(1)]
  POI Distribution: The POI distribution based method is to directly apply $K$-means and CRF clustering methods on the POI distribution features.
  \item[(2)]
  Collaborative POI: In addition, to considering latent structure and avoiding missing data in TF-IDF based Bag-of-POIs, collaborative filtering methods have been used to obtain latent semantic representation of POIs \cite{yuan2015discovering}
  . In our comparison, collaborative POI is obtained by SVD decomposition and represented by the eigenvecotrs that are associated with the top $t$-th maximum eigenvalues. Then, $K$-means and CRF clustering methods are respectively employed based on collaborative POI features.
  \item[(3)] Topic Modeling:  {By regarding the regions as a bunch of documents and the POI is the words of each document, then the region functionalities actually play the same of the topics in the documents. As the topics determine the words generation in each documents, the region functionality also determines the POI generation in each region. In addition, the human activities are the prior knowledge of the generating distribution of the POIs, as the human activities would also influence the POIs. Based on the each topic of the documents, the documents can be clustered in the similar topic clusters. In the same way, region functional zones can be clustered by those region functionalities that computed as the topics. As a typical topic model, LDA, therefore, is used to discover the region functionalities}
  based on the HAP and then cluster functional zones by $K$-means. For LDA model, we choose 10 topics with 1000 iterations, following the settings in \cite{yuan2018discovering}.
\end{itemize}

In our experiment, we learn our latent region representation by Algorithm \ref{alg1}. Two clustering approaches: $K$-means and CRF are utilized on our latent region representations to cluster $S$ different functional zones. We also cluster our latent semantic HAP by just using $K$-means to compare with Topic Modeling method \cite{yuan2018discovering}.

To verify our results, we compare our results with land use map of Raleigh and check whether most famous regions are assigned into their proper functional zones.

\subsection{Results}
\subsubsection{Functional Zones Comparison}

\begin{figure*}[!ht]
	\centering
	\subfloat[SVD of POI + K-means]{\includegraphics[width=0.33\textwidth]{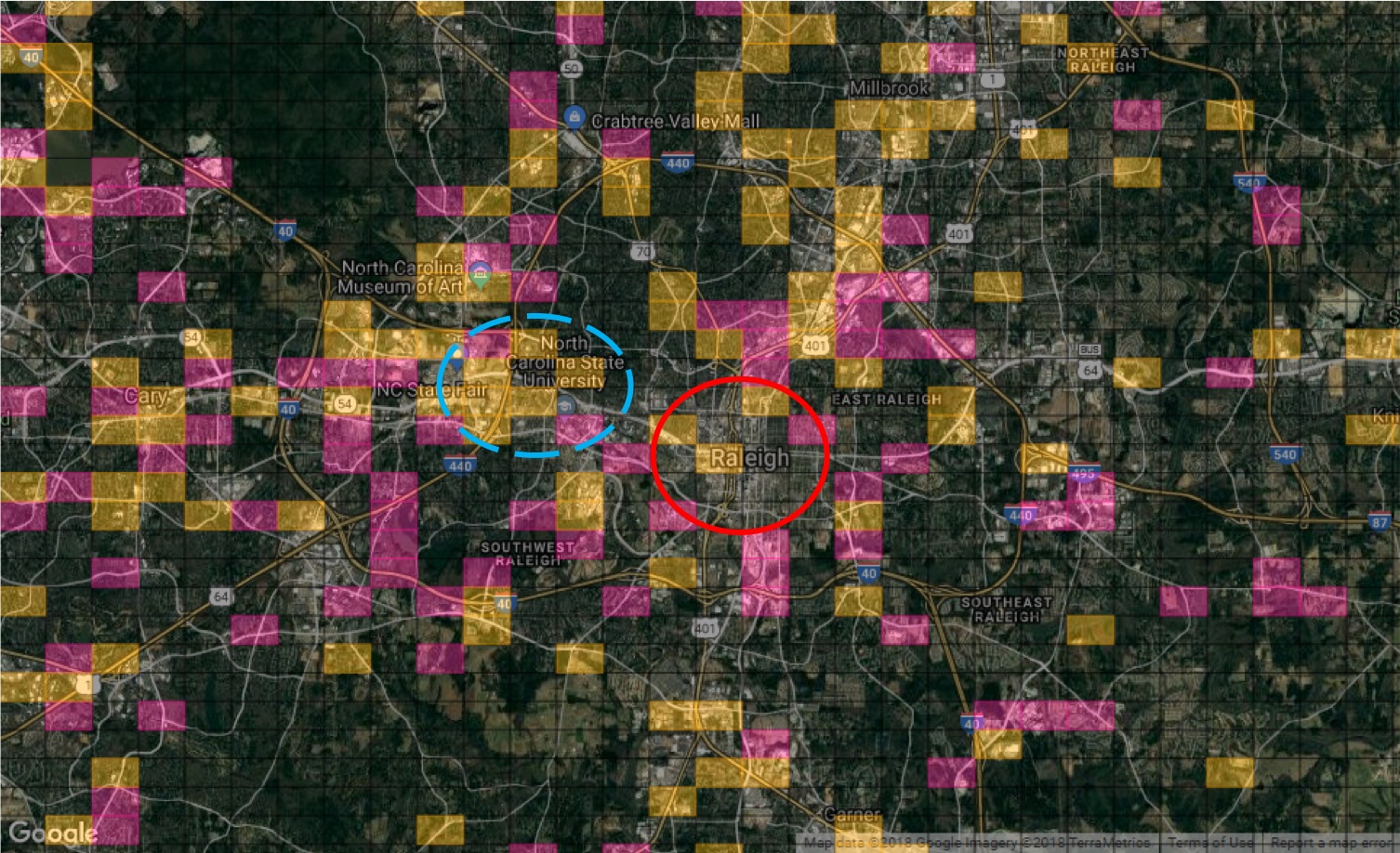}\label{fig:copoi-kmeans}}
	\hfill
	\subfloat[POI + K-means]{\includegraphics[width=0.33\textwidth]{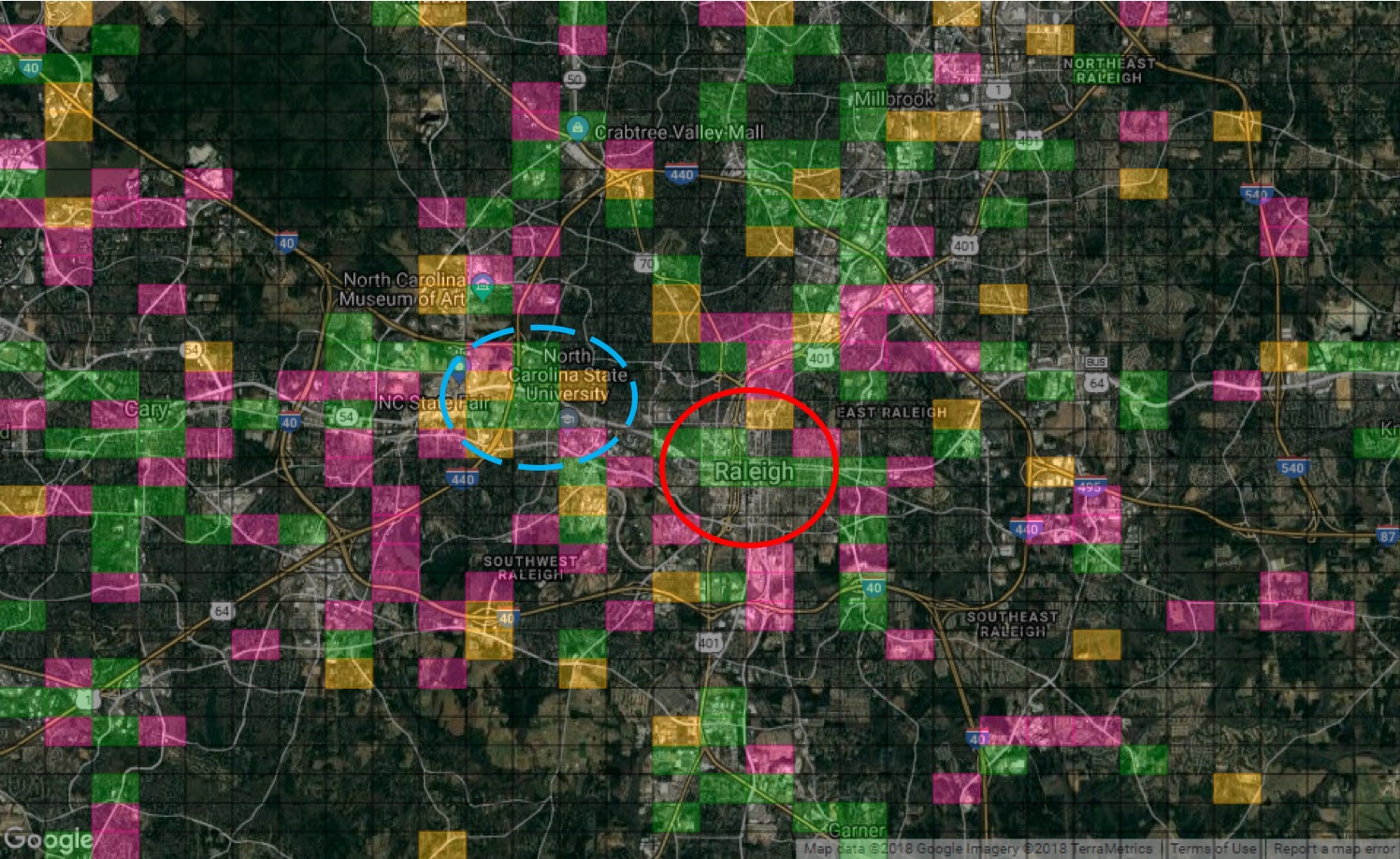}\label{fig:poi-kmeans}}
	\hfill
	\subfloat[Our Latent Representation + K-means]{\includegraphics[width=0.33\textwidth]{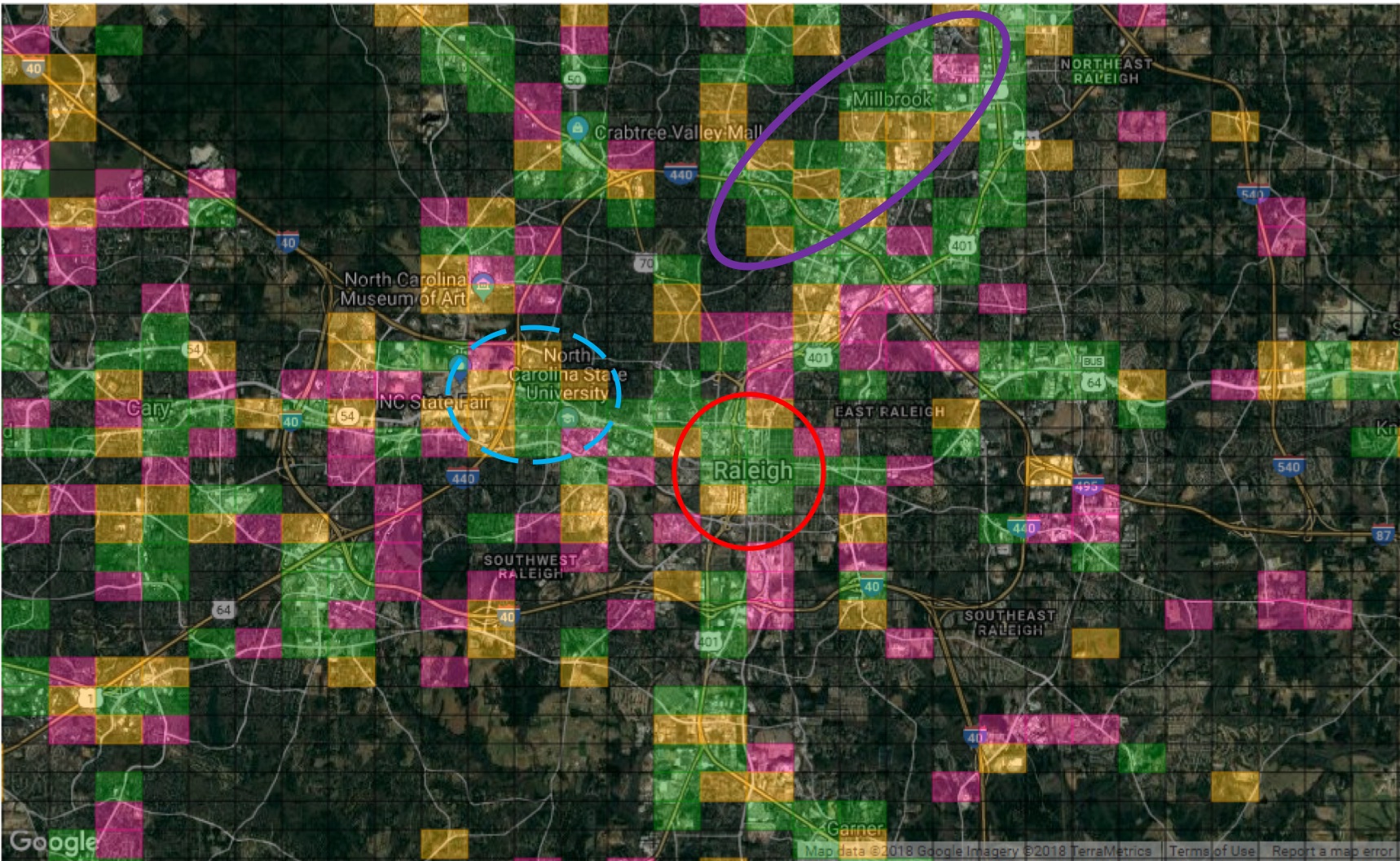}\label{fig:our-kmeans}}
	\vfill
	\subfloat[SVD of POI + CRF]{\includegraphics[width=0.33\textwidth]{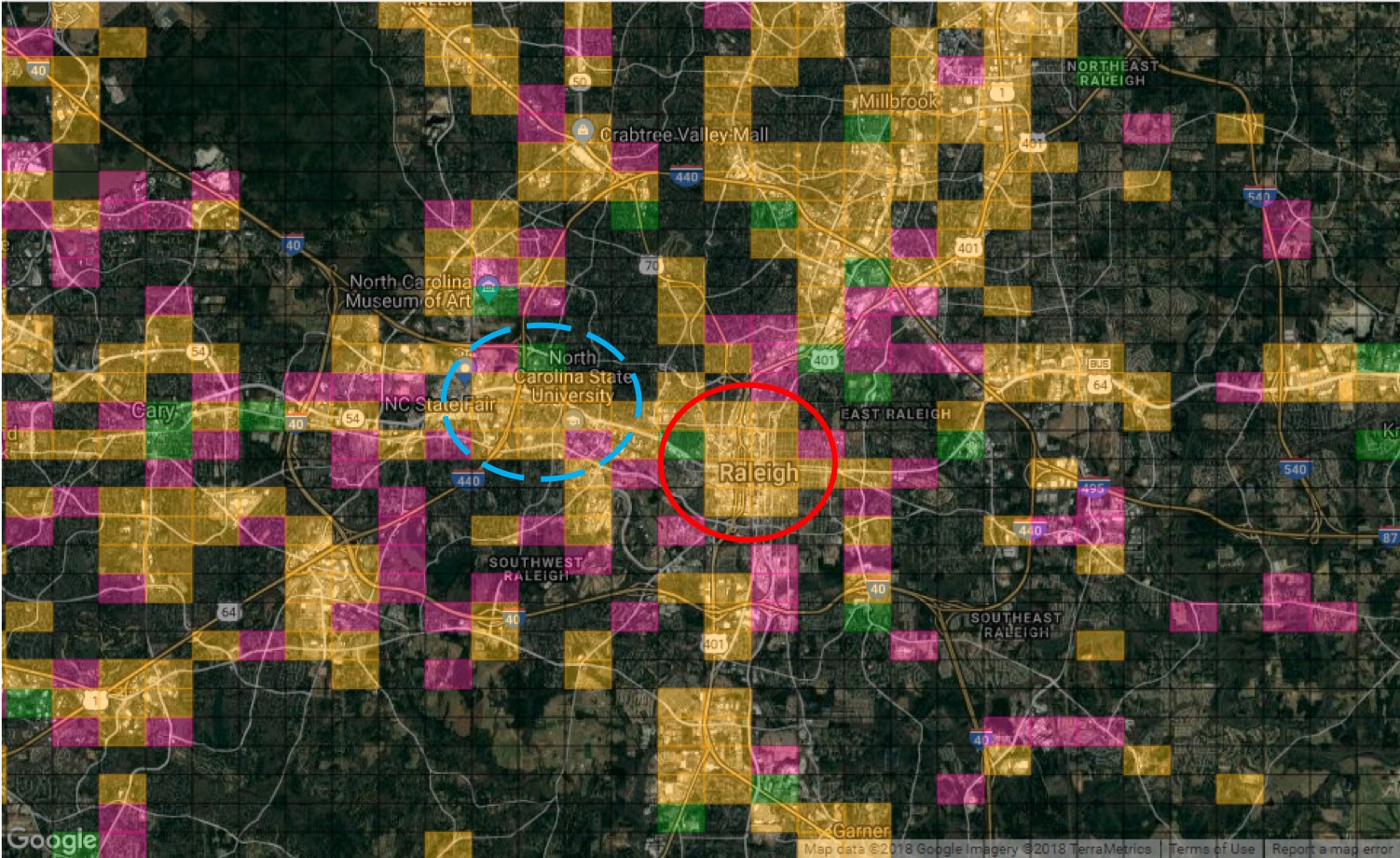}\label{fig:copoi-crf}}
	\hfill
	\subfloat[POI + CRF]{\includegraphics[width=0.33\textwidth]{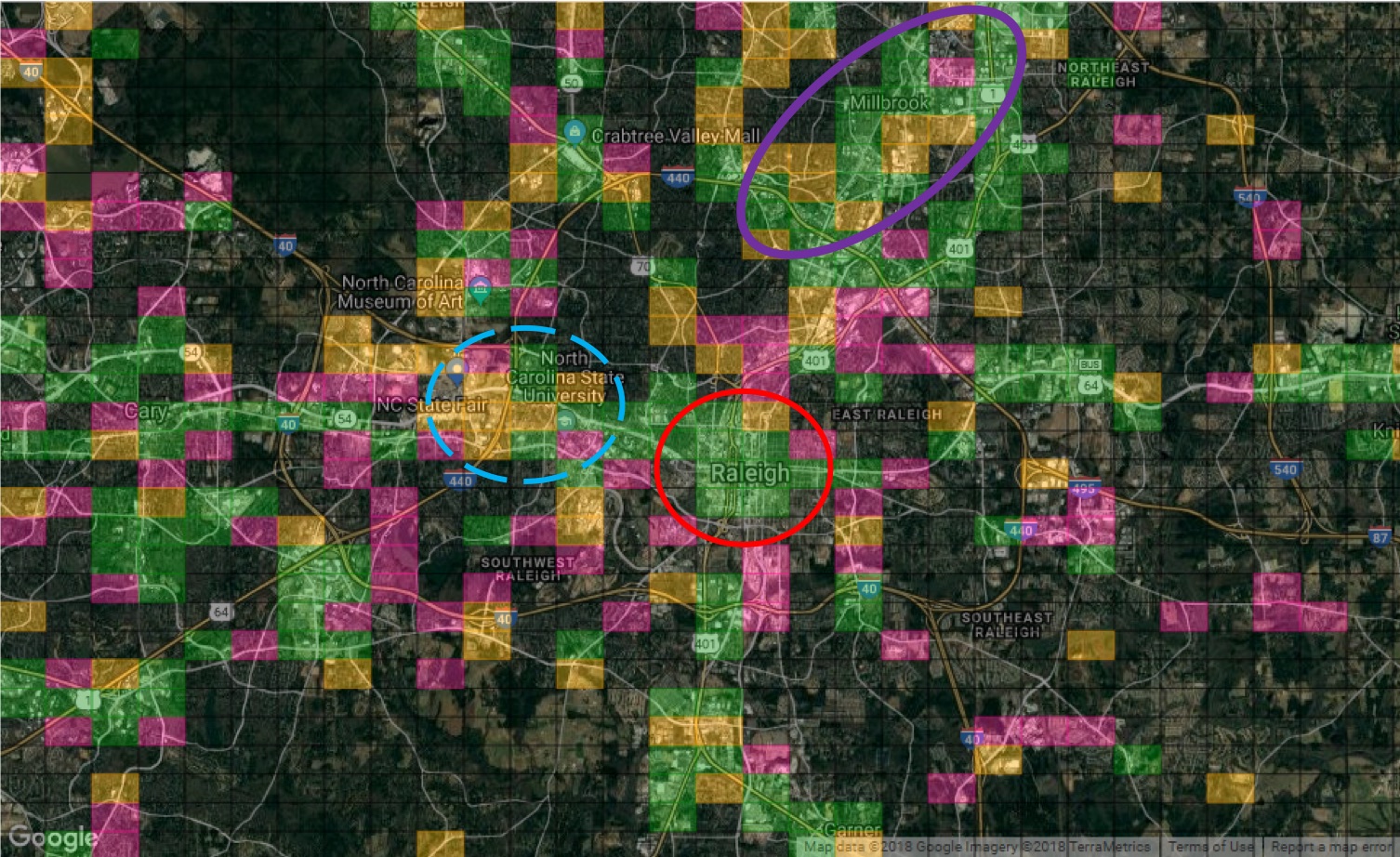}\label{fig:poi-crf}}
	\hfill
	\subfloat[Our Latent Representation + CRF]{\includegraphics[width=0.33\textwidth]{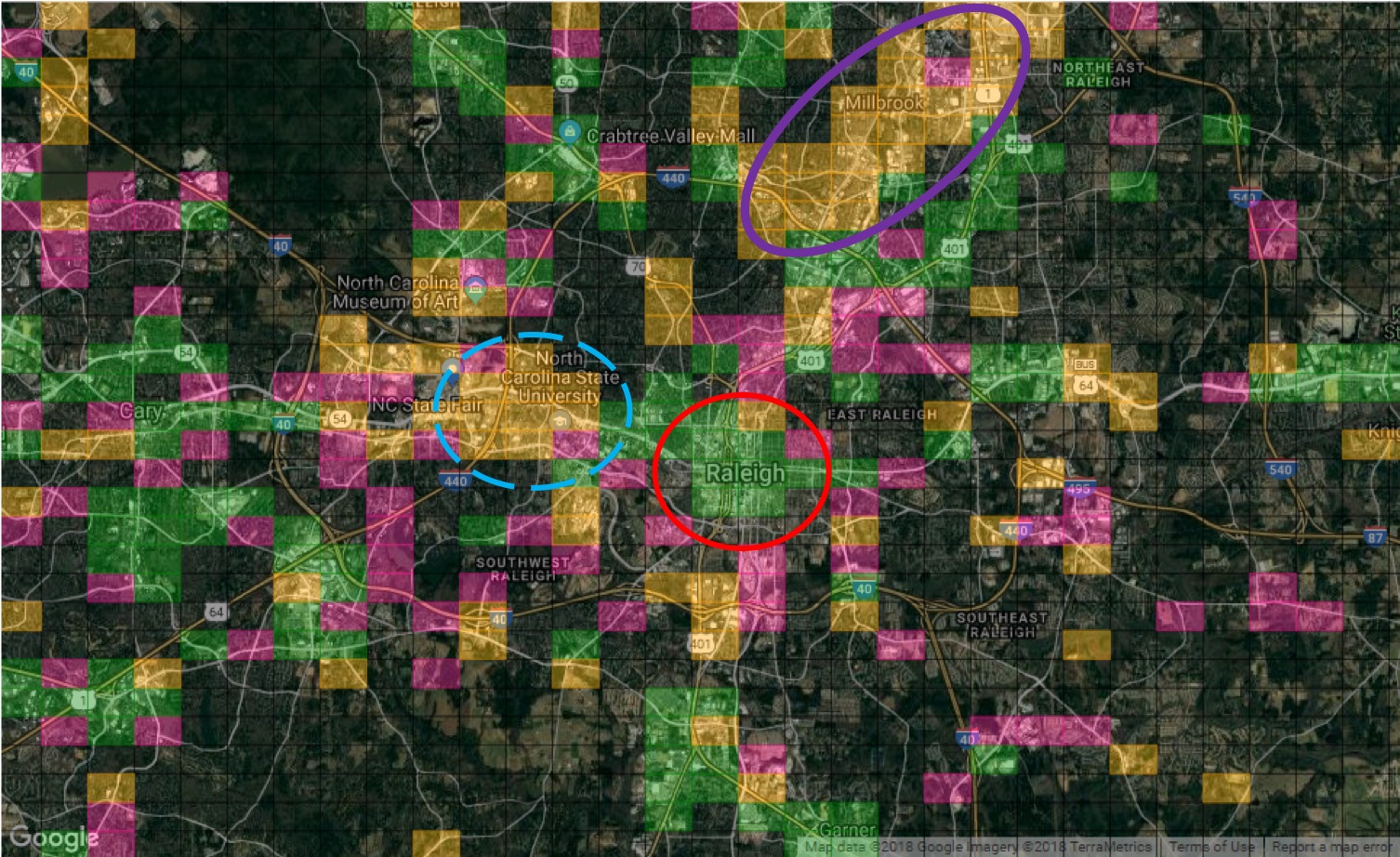}\label{fig:our-crf}}
	\vfill
	\subfloat[Our Latent Representation + CRF for the Whole Map Area]{\includegraphics[width=0.33\textwidth]{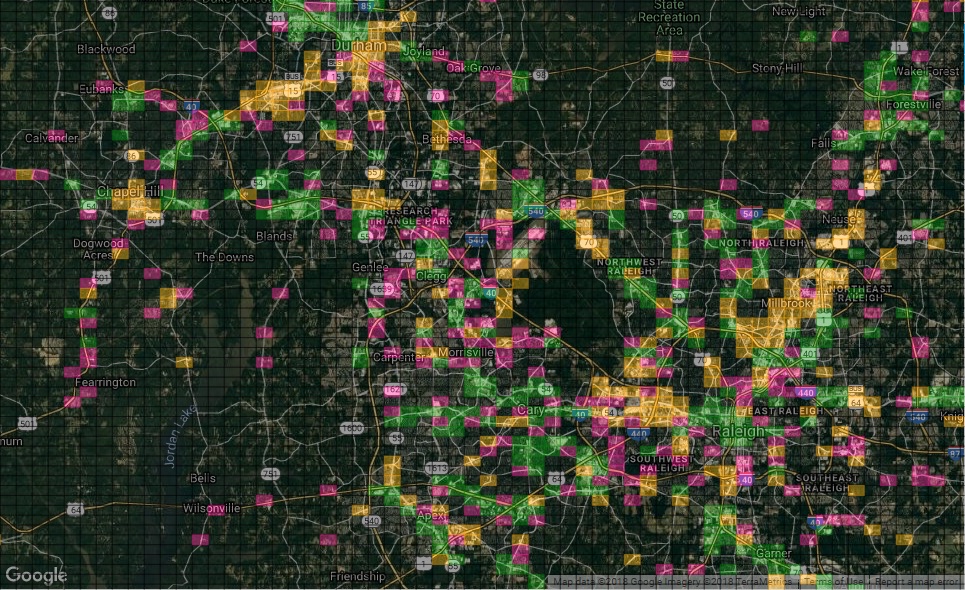}\label{fig:our-kmeans-whole}}
	\hfill
	\subfloat[LDA topics + K-means]{\includegraphics[width=0.33\textwidth]{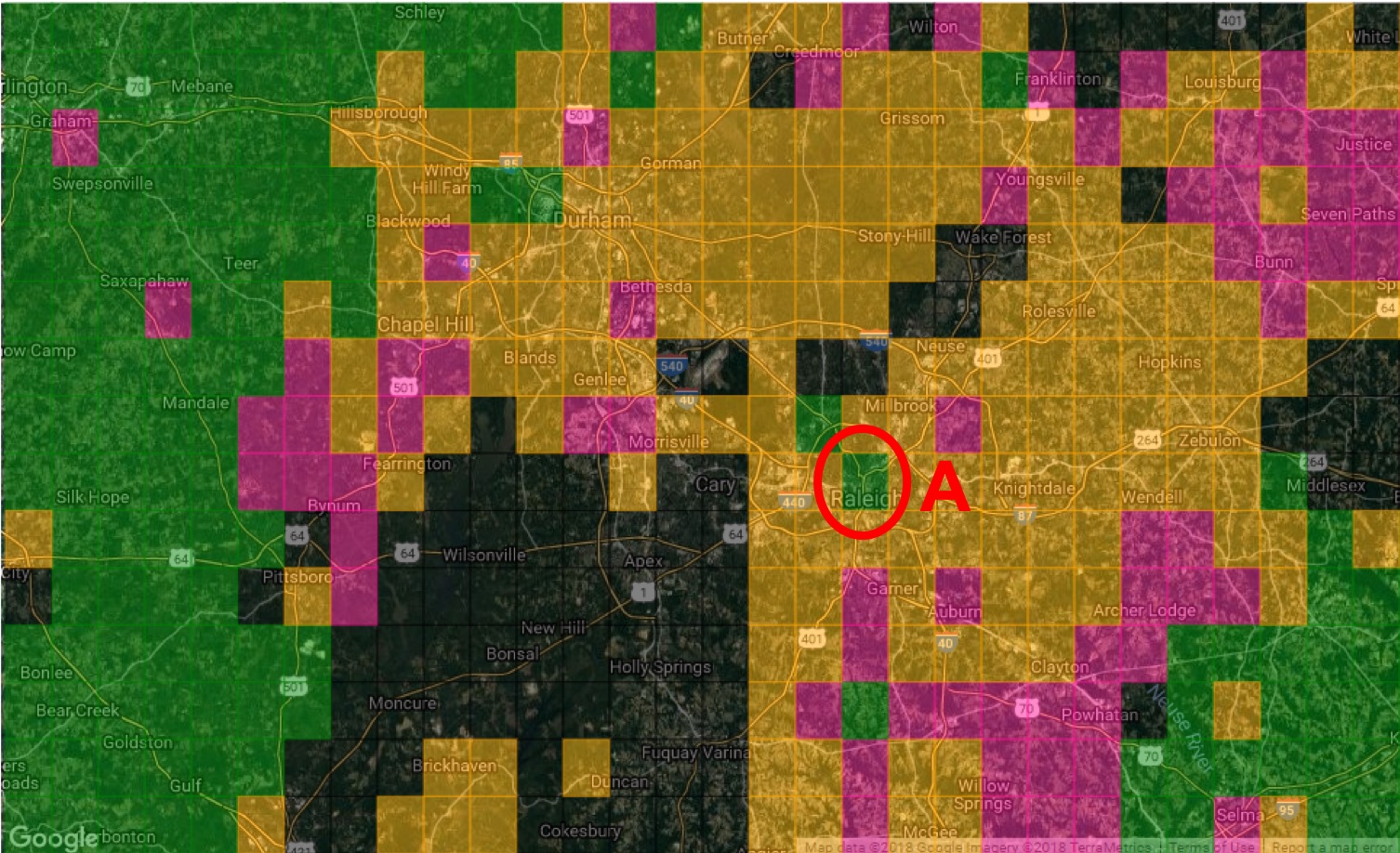}\label{fig:topic-kmeans}}
	\hfill
	\subfloat[Our semantic HAP + K -means]{\includegraphics[width=0.33\textwidth]{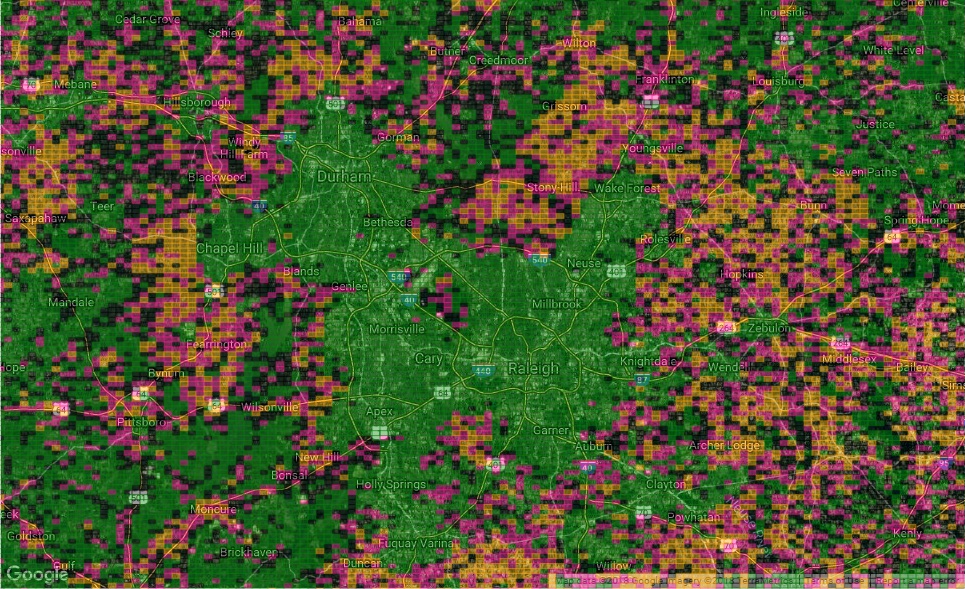}\label{fig:latentZ-kmeans}}
	\caption{Comparison of different region representations.}\label{fig:mapcomparsion}
\end{figure*}

We aggregate 4 functional zones by hereabove methods for Raleigh city  and indicated by different colors, as shown in Fig. \ref{fig:our-kmeans-whole}. In order to better compare the different methods, we zoom in on the results of major parts of Raleigh and demonstrate them in Fig. \ref{fig:copoi-kmeans}-\ref{fig:our-crf}. It is noting that different colors in different figures may indicate different functional zones. The results shown in  Fig.\ref{fig:copoi-kmeans}, \ref{fig:poi-kmeans}, \ref{fig:copoi-crf} and \ref{fig:poi-crf} are only based on clustering of the POI information, while the results shown in Fig.\ref{fig:our-kmeans} and \ref{fig:our-crf} are based on clustering the fusion features of both POI and HAP information. Fig. \ref{fig:copoi-kmeans}-\ref{fig:our-kmeans} are all based on $K$-means clustering, while Fig. \ref{fig:copoi-crf}-\ref{fig:our-crf} are all based on CRF clustering. Fig. \ref{fig:topic-kmeans} and \ref{fig:latentZ-kmeans} in the third row figures are based on the clustering of only the HAP information.

\noindent\textit{A. Quantitative Comparison}

 {To quantitative evaluate our framework, the information retention and the unbiased degree are introduced as two direction measurements. As the dataset is extremely sparse, to avoid loss such precious information, the information retention is used to evaluate how much original dataset information is retained after the clustering. In another hand, since POIs are biased and only contain the commercial POIs, then most regions are easily clustered into the same clusters when their functionalities may totally different. Therefore, to measure how much information balanced in our dataset to avoid the bias of POIs, the unbiased degree is included to measure the how balanced the clusters are.}

\begin{table}[htp]
    \centering
    \resizebox{0.99\columnwidth}{!}{
    \begin{tabular}{|c|c|c|c|c|}
    \hline
        &our feature &  our feature   & Collaborative POI  & POI \\
        &w. CRF &w. K-means  &w. K-means & w. K-means\\
    \hline
    Before Clustering &     7.52\%& 7.52\% & 7.52\% & 7.52\% \\
    After Clustering&     7.52\%& 7.52\% & 5.27\% & 5.97\% \\
    \hline
    Information Retention&     \underline{\textbf{100\%}}& \underline{\textbf{100\%}} & 70.08\% & 79.39\% \\
    \hline
    \end{tabular}
    }
    \caption{Information Retention}
    \label{tab:information}
\end{table}

Like second and third-tier cities, Raleigh contains very few POIs. There are only 7.52\% regions with POI information in our experiments, and the sparsity of POI matrix is up to 99.11\%. The black clusters in Fig. \ref{fig:copoi-kmeans}-\ref{fig:our-crf} represent missing the POI information group. Since this group is really large, other undistinguished features are easily clustered into this missing POI information group. As shown in Fig. \ref{fig:poi-kmeans}, only using POI distributions is not discriminative enough, many regions are regarded similar to the missing POI information one.  {The detailed information retention are listed in Table \ref{tab:information}.}

For the benchmark (1), after directly using k-means clustering on POI distribution, the regions containing POI information significantly decreases from 7.52\% to 5.97\%, which misses 20.61\% information. In Fig. \ref{fig:poi-kmeans}, it shows that only the top parts of downtown Raleigh that circled in red are not black and carrying POI information.

For the benchmark (2), in order to reduce the sparsity and increase the latent semantic meanings of POI information, SVD is then applied on POI matrix. However, with limited data in POI matrix, SVD decomposition can not explore more latent structure. The result in Fig. \ref{fig:copoi-kmeans} is even worse than just using the POI distribution, i.e., the benchmark (1). It also shows that most regions are clustered into 3 groups (the black, yellow and pink ones), when 4 groups should be clustered. Additionally, only 5.27\% regions contain the POI information, while 5.97\% regions preserve the POI information for the benchmark (1). This loses 29.92\% of all the information, which is mainly caused by a lack of discriminative features in Collaborative POI.

In Fig. \ref{fig:our-kmeans}, our latent region representations are fused both POI and HAP information. It shows that such latent region representations are distinguished enough to preserve all of the 7.52\% POI information regions, \textit{i.e.}, 100\% information is preserved. Most of the regions with similar functionalities are in the same clusters. For example, the major parts of the regions in the red circle are assigned a green color, which actually is the downtown area of Raleigh. And half of the regions in the blue circle are assigned an orange color, where the university area should be located. Our latent region representation is more disciminative than the other benchmarks.

 {To preform the unbiased degree of the clusters,} $Unbiased$ $Degree = 1- silhouette~score,$  {where silhouette score proposed in \cite{rousseeuw1987silhouettes} is a measure of the cluster compactness.In contract to the general case that the lower decentralization degree, the worse, in our case, the higher decentralization degree, the better. Because that means biased POI are balanced when the knowledge is propagating between the unbiased intrinsic interaction between POIs and HAPs.}
\begin{table}[htp]
    \centering
    \resizebox{0.99\columnwidth}{!}{
    \begin{tabular}{|c|c|c|c|}
    \hline
        our feature &  our feature   & Collaborative POI  & POI \\
        w. CRF &w. K-means  &w. K-means & w. K-means\\
    \hline
         \underline{\textbf{6.01\%}} & \textbf{5.73\%} & 2.61\% & 2.28\% \\
    \hline
    \end{tabular}
    }
    \caption{Unbiased Degree}
    \label{tab:silhouette_score}
\end{table}

 {As shown in Table \ref{tab:silhouette_score}, using the simple $K$-means clustering on the distribution of POI obtains the lowest unbiased degree, because of the dominance of the missing POI regions. When the Collaborative POI feature is used, the decentralization degree is slightly increased 0.33\% because its SVD decomposition seeks the latent representation of POIs and recovers the unbiased representations to some extent. However, when using our proposed feature even for $K$-means clustering, the decentralization degree is significantly increased by 3.12\%, which is 2 times more than Collaborative POI features. That is, our proposed method did learn the unbiased information from the intrinsic structure of POIs and HAPs. And later qualitative comparison also show that our balanced region representations do reveal the true region functionalities. Moreover, when using CRF for clustering, the functionality consistence is also balanced among fine-grained regions, which results in another 0.28\% more increasing of the unbiased degree than the one of $K$-means clustering. That is, 2.3 times more than the result of the Collaborative POI and 2.6 times more than the result of POI.

Based the quantitative comparison, our proposed framework not only retains the 100\% information of the data, but also balance the region representations and clusters for true functional zone discovery.}

\noindent\textit{B. Qualitative Comparison}

 {In addition to the quantitative comparison, we also show the functional zones on the map to compare the qualities of discovering the functional zones. The land-use map is also used for more accurate comparison.}

Comparing with the results of $K$-means shown in Fig. \ref{fig:copoi-kmeans}-\ref{fig:our-kmeans}, the results of CRF in Fig. \ref{fig:copoi-crf}-\ref{fig:our-crf} are more consistent. Most of the regions are assigned the same colors. The blocks of regions in the same color are more numerous than those of using $K$-means.

For the benchmark (2), although the CRF result based on Collaborative POI in Fig. \ref{fig:copoi-crf} is much better than $K$-means one, most of the regions are in the same functional zones, which lose its meanings to discover the functional zones. For example, the functionality of the downtown part that is circled in red should be different from the university area that is circled in blue.

\begin{figure}[!htb]
	\centering
	\subfloat{\includegraphics[width=0.4\textwidth]{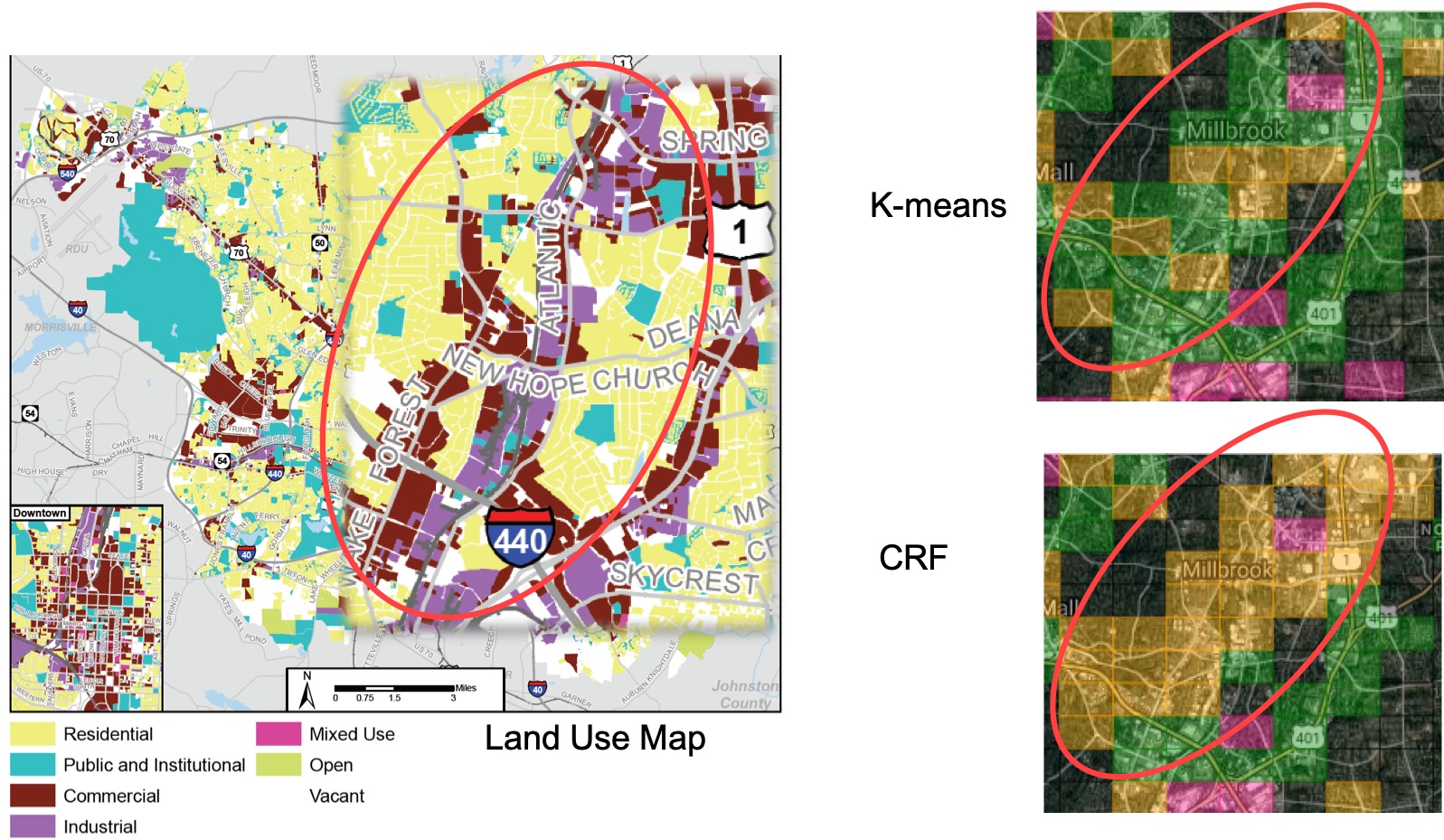}\label{fig:440-compare}}
	\vfill
	\subfloat{\includegraphics[width=0.4\textwidth]{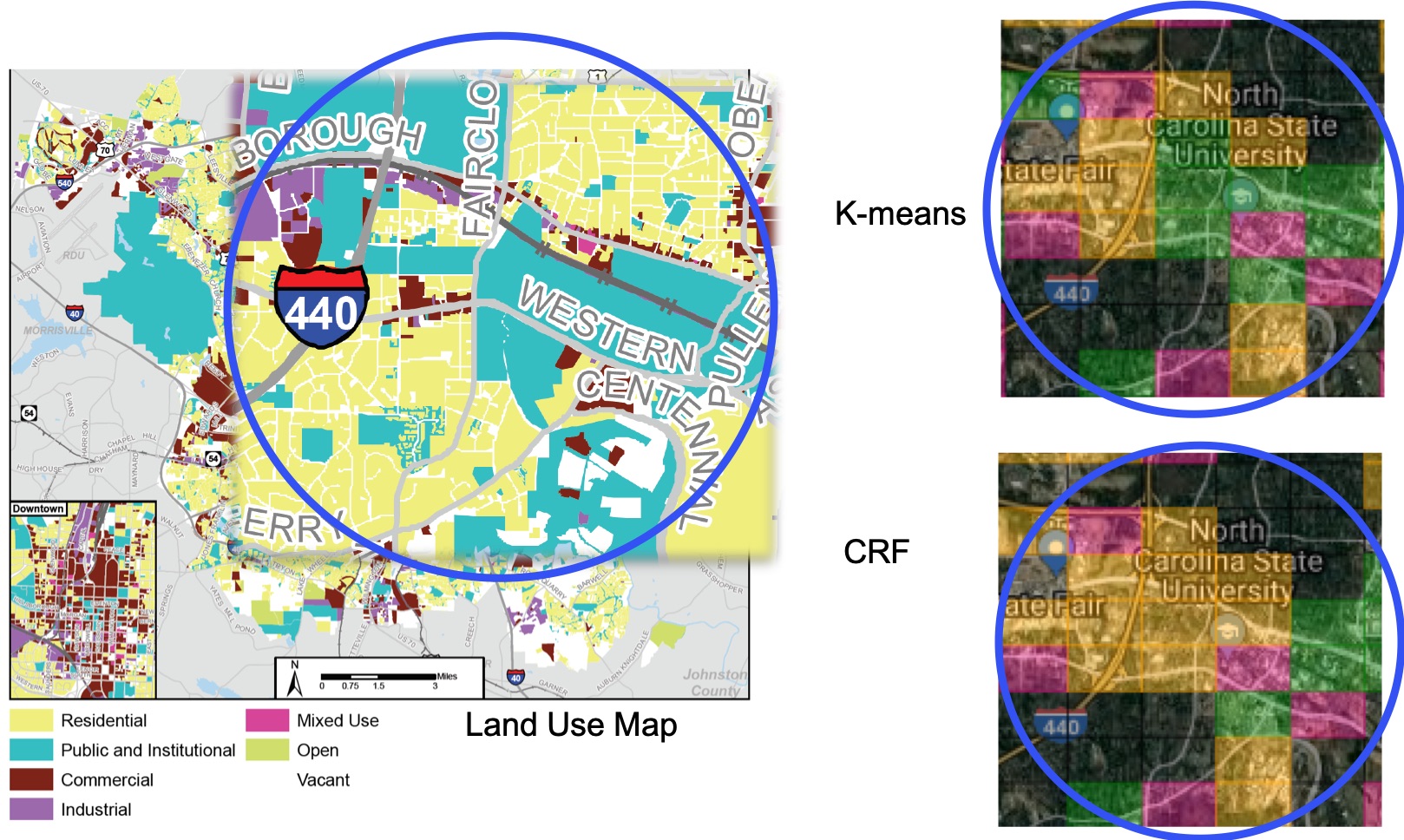}\label{fig:ncsu-compare}}
	\caption{Comparison of different clustering methods based on Latent Region Representation.}
	\label{fig:areacomparsion}
\end{figure}

For the benchmark (1), the CRF result is based on POI distributions. As shown in Fig. \ref{fig:poi-crf}, it has a more uniform clustering result than benchmark (2), which is more meaningful. But it is still not so consistent as the results of our latent region representation shown in Fig. \ref{fig:our-crf}. In Fig. \ref{fig:poi-crf}, parts of the region functionalities in the university area circled as blue are the same as those in the downtown area circled as red. Also, there are many industries located in the purple circle, which should have similar functionality as the university area, especially for workdays. However these regions in the purple circle are assigned half green and half orange in Fig. \ref{fig:poi-crf}. To look at the regions in both the blue and purple circles in Fig. \ref{fig:our-crf}, they are almost assigned to orange color. Therefore, our latent region representation is still discrimative than other benchmarks. Additionally, the CRF clustering of our latent region representations has a more consistent result than the $K$-means clustering does as shown in Fig. \ref{fig:our-kmeans} and \ref{fig:our-crf}. A more detailed comparison of land-use map and the results of both $K$-means and CRF are shown in Fig. \ref{fig:areacomparsion}.

\begin{table*}[htp]
	\centering
	\resizebox{1.7\columnwidth}{!}{
	\begin{tabular}{|c|c|c|c|c|c|c|}
		\hline
		Ranked &Green &$G_1$ & Orange &$G_2$ & Pink &$G_3$\\
		\hline
		1&fast food & 0.538 &financial service&0.453 & shipping &1.172\\
		2&financial service &0.453 &fast food&0.352 & fuel&0.871\\
		3&coffee bar & 0.308& auto service&0.208 &tutoring&0.110\\
		4&auto service&0.259 &auto supply&0.207 & hotel&0.097\\
		5&eateries&0.257 &pharmacy &0.114 &grocery&0.087\\
		6& pharmacy&0.232 &eateries&0.095 &home improvement&0.082\\
		7&pet&0.195 &convenience&0.092 &auto rental&0.010\\
		8& personal care&0.173& auto dealers&0.055& park/lake(camping site)&0.004\\
		9&home improvement&0.099 &grocery&0.041 &machinery&0.004\\
		10&shopping center&0.089 &coffee bar&0.035 &-&-\\
		\hline
	\end{tabular}
	}
	\caption{Map Color and Its Significant POIs with Ranked Difference $G_s$}
	\label{tab:importantpoi}
\end{table*}

For the benchmark (3), since \cite{yuan2018discovering} demonstrated that LDA model has a similar result of functional zones as the one of DMR model, and DMR model which takes the POI information into account only change few areas but spending more time and memory for training, it is fair to choose LDA to be the benchmark for comparison. As constrained memory limitation for LDA training, we use Geohash level $5$ $4900~m~ \times ~4900~ m$ grids to segment the same city map and show its results in Fig. \ref{fig:topic-kmeans}. As shown in Fig. \ref{fig:topic-kmeans}, the result of LDA nicely indicates that the map is segmented into 4 different parts based on the spatial locations. However, there are still some not well clustered regions. For example, the green part that is circled in rad and labeled with "A" is the downtown area of Raleigh, and it should be a high-level-activity area. However, the activity level in other green areas in Fig. \ref{fig:topic-kmeans} is much lower than the ones of downtown Raleigh. That is, downtown Raleigh should not be assigned to the same group as other green areas are. Also, the orange area covers for the most part cities, while many vacant and nature places are also included in this orange area. For a fair comparison, we also used $K$-means to cluster our semantic HAP and show its results in Fig. \ref{fig:latentZ-kmeans}. In Fig.  \ref{fig:latentZ-kmeans}, the centre green areas are the major cities, such as, Raleigh, Durham, Chapel-Hill, Cary and Apex. The orange areas represent a middle-level activity area, and the pink areas denote a low-level-activity area. Our latent transition pattern $Z$ is capable highlight how urban these areas are, while LDA still blurs the high-level-acitivity area with other level-activity areas. It worth nothing that, our proposed method is based on the Geohash level $6$ (1 region unit of Geohash level $5$ covers 32 region units of Geohash level $6$), which has a finer granularity than LDA does.

Overall, using the CRF to cluster our latent region representations that fused both POI and HAP information achieves a superior performance than other approaches. The latent region representations are more disriminative. The CRF method achieves a more consistent clustering result. And our proposed framework supports a finer granularity with a practical implementation time.

\subsubsection{Functional Zones Annotation}
According to our functionality estimation, we calculate the normalized POI distribution $NPR$ for each different functional zone. The difference $G_s$ of each functional zone is then computed, and the components of $G_s>0$ are listed by decreasing the significance in Table. \ref{tab:importantpoi}. Since areas and cities in our experiment are not modern cities, they do not have a complex city structure. We thus consider 4 different functional zones in our experiments. The area annotations are based on the results of clustering our latent region representations by CRF, as shown in Fig. \ref{fig:our-crf} or \ref{fig:our-kmeans-whole}. As the black areas are where no POI information regions are, and most regions in black are residential, vacant and natural areas, we did not list them in Table. \ref{tab:importantpoi}. In terms of the ranked difference of significant POIs in the table, we annotate the regions as follows:

\noindent\textbf{Commercial/Downtown/Entertainment Area (Green regions)}: The distinguished POI in this functional zone are Pet, Personal Care and Shopping center, which are much significant in this area than others. Pets, Personal Care and Shopping center all provide service and merchandise, which indicates this functional zone is probably a commercial area. The other highly ranked POIs in this area also follow the characteristics of the commercial area, such as, financial service and coffee bar. Additionally, the famous shopping malls, such as Southpoint and Crabtree are all located in this functional zone.

\noindent\textbf{Working/Education Area Mixed with Commercial Area in Inner City (Orange regions)}: Since we only count for the workdays, this area is annotated as a mixed use area. During the working time, its function is working/education. During other times, its commercial area. The $G_s$ of this area focus on the financial service, fast food, and auto service, which is easy to compose a commercial area. Meanwhile, looking at the Fig. \ref{fig:areacomparsion}, both industrial and Public and institutions are mixed with commercial part and located in orange regions, such as NC state, Duke, Chapel-Hill and Wake tech Universities.

\noindent\textbf{Working/Education Area in Outer City (Pink regions)}: Most significant POIs in this functional zone are shipping, fuel, hotel and auto rental, which all refers to a outer-side of a city. Additionally, as our dataset is based on the workdays, and some industrial and tutoring are also located in the outer city, so this area is implied to be a  working/education area. Consequently, this functional zone is annotated as a working/Education area in outer City. Since this area include few commercial area, it is different from above mixed with commercial one. According to Fig. \ref{fig:our-kmeans-whole}, it is easy to find pink area is in outer side of centralized area. Moreover, the Research Triangle Park (RTP), a large company area is also assigned in the pink area, where it locates in middle part among the triangle part that consisted by the outer side of the cities of Raleigh, Durham and Chapel Hill.

\section{Discussion}\label{sec:discussion}
 {For further discussion, in this section we introduce the proper conditions to use our method. In our framework, there are two presumptions to propose the whole pipeline. One is the fine-grained region units and their neighbors may share the same region functionalities. Another is the HAP and POI pattern are follow the same introduced intrinsic pattern. For coarse grained granularity, the neighbors of each region unit are not in the same clusters, and each of them has their independent functionalities. Then CRF clustering for the spatial location consistency will be harmful for the final clustering results rather than helpful. Besides, the introduced inner pattern between POI and HAP is transferred from \cite{wang2017human}, whose knowledge was discovered on the traffic pattern in Beijing, China. Other city may not share the same pattern of Beijing, then the latent region presentation is not accurate for clustering any more. In our case study, we did not have any prior knowledge of Raleigh city, thus we transfer the knowledge of Beijing city as a cold start, which also performs a good result. But if this pattern are totally different, the clustering result will be affected. Since in our framework, we have a mask to deal with the sparse data, therefore it also works for dense dataset, which should not be a concern.}

 {Consequently, the proposed methodology would be more favorable for fine-grained granularity segmentation of regions. Also, our framework is a really good starting when lacking of the knowledge of the pattern in the dataset or obtaining a similar pattern between the HAP and POI.}

\section{Conclusion} \label{sec:conclusion}
In this paper, we proposed a framework for discovering functional zones on the basis of a novel algorithm to learn discriminative latent representations of regions and spatial clustering method. The proposed new algorithm leverages the latent effects between POIs and HAP, and successfully fuses the information of highly sparse POIs and HAP, and simultaneously enhances the discriminiatve ability. An intrinsic knowledge that explored from the real data are transferred in our framework to cope with biased data. Moreover, we also utilize the label consistency of CRF to preserve the spatial information when aggregating the functional zones, which unites the clustering of fine-girded regions. The difference of normalized POI distributions is employed to estimate the functionalities for a better annotation. Comparing with LDA-based methods, our method produces better results and a finer granularity with practical implementation conditions. The proposed framework can efficiently discover different functional zones and help designers to fully understand the urban structure.

\section{Acknowledgement}
We would like to thank Valassis Digital to support this research and provide access to their data. We would like to also show our gratitude to Schaun Wheeler, Mark Lowe and Jeff Hicks for sharing their valuable inputs and suggestions during the course of this research.

\printcredits

\bibliographystyle{cas-model2-names}

\bibliography{eg}

\end{document}